%% file: Combined.tex
\documentclass[aps,pre,twocolumn,floatfix,superscriptaddress]{revtex4-1}
\usepackage{mathtools}
\usepackage{amsmath}  
\usepackage{amsfonts} 
\usepackage{graphicx} 
\usepackage{epstopdf}
\usepackage{gensymb}
\usepackage{physics}
\usepackage{xcolor}
\usepackage{siunitx}

\newcommand{\CC}{\mathbb{C}}
\newcommand{\Cb}{\mathbf{C}}
\newcommand{\Mb}{\mathbf{M}}
\newcommand{\Hb}{\mathbf{H}}

\newcommand{\PP}{\mathcal{P}} 
\newcommand{\DD}{\boldsymbol{\mathcal{D}}}
\newcommand{\dD}{\boldsymbol{D}} 
\newcommand{\Psib}{\boldsymbol{\Psi}}

\begin{document}
\title{Realization of active metamaterials with odd micropolar elasticity}

\author{Yangyang Chen}
\thanks{These authors contributed equally to this work.}
\affiliation{Department of Mechanical and Aerospace Engineering, University of Missouri, Columbia, MO, 65211, USA}

\author{Xiaopeng Li}
\thanks{These authors contributed equally to this work.}
\affiliation{Department of Mechanical and Aerospace Engineering, University of Missouri, Columbia, MO, 65211, USA}

\author{Colin Scheibner}
\thanks{These authors contributed equally to this work.}
\affiliation{James Franck Institute, The University of Chicago, Chicago, IL, 60637, USA}
\affiliation{Department of Physics, The University of Chicago, Chicago, IL, 60637, USA}

\author{Vincenzo Vitelli}
\email{vitelli@uchicago.edu}
\affiliation{James Franck Institute, The University of Chicago, Chicago, IL, 60637, USA}
\affiliation{Department of Physics, The University of Chicago, Chicago, IL, 60637, USA}
\affiliation{Kadanoff Center for Theoretical Physics, The University of Chicago, Chicago, IL, 60637, USA}

\author{Guoliang Huang}
\email{huangg@missouri.edu}
\affiliation{Department of Mechanical and Aerospace Engineering, University of Missouri, Columbia, MO, 65211, USA}

\begin{abstract}
Materials made from active, living, or robotic components can display emergent properties arising from local sensing and computation.
Here, we realize a freestanding active metabeam with piezoelectric elements and electronic feed-forward control that gives rise to an odd micropolar elasticity absent in energy-conserving media. The non-reciprocal odd modulus enables bending and shearing cycles that convert electrical energy into mechanical work, and vice versa. The sign of this elastic modulus is linked to a non-Hermitian topological index that determines the localization of vibrational modes to sample boundaries. At finite frequency, we can also tune the phase angle of the active modulus to produce a direction-dependent bending modulus and control non-Hermitian vibrational properties. Our continuum approach, built on symmetries and conservation laws, could be exploited to design others systems such as synthetic biofilaments and membranes with feed-forward control loops. 
\end{abstract}

\maketitle

\begin{figure*}[ht!]
\centering
\includegraphics{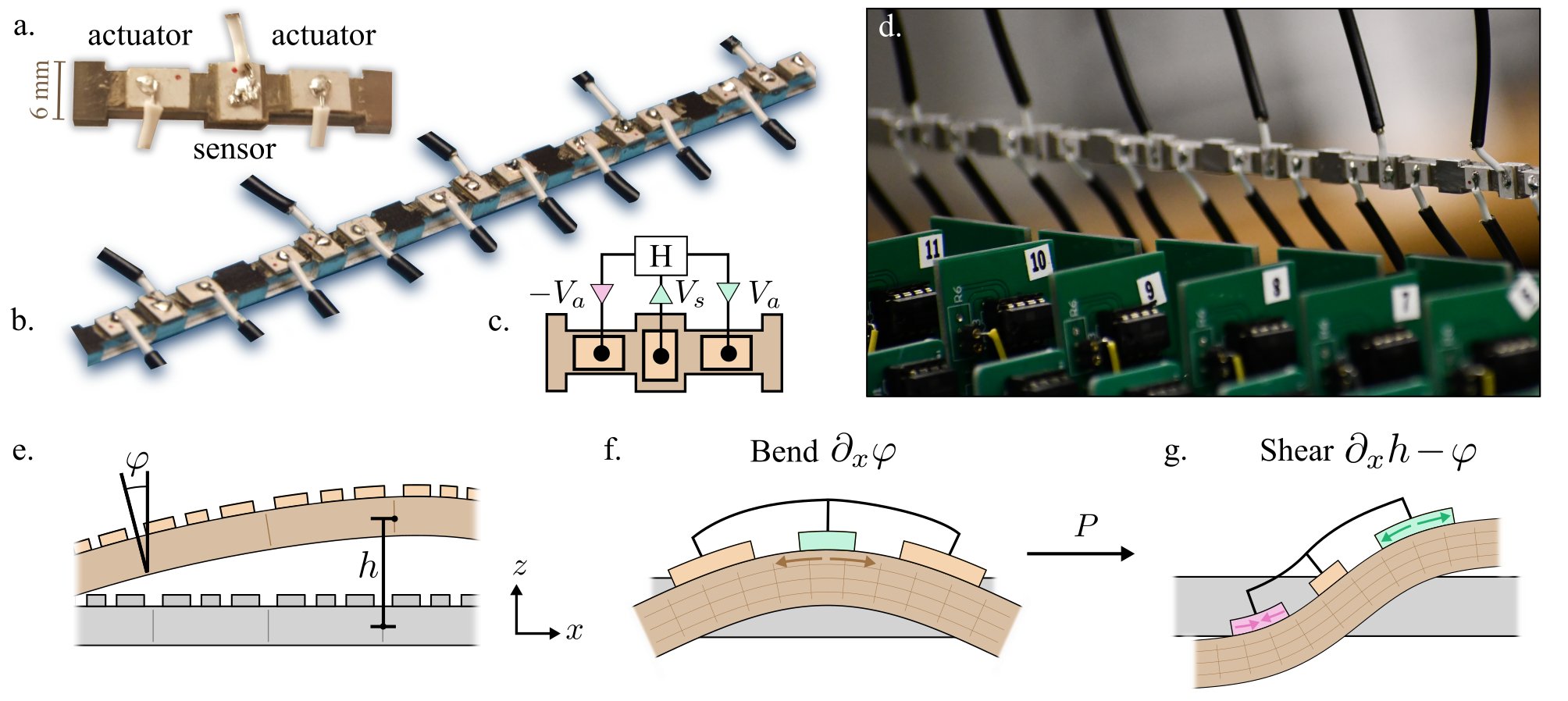}
\caption{ {\bf Design and mechanics of an odd micropolar metabeam.} {\bf (a)}~A single unit cell featuring three piezoelectric patches mounted on a beam: one that acts as a sensor, and two that act as actuators. 
{\bf (b)}~A segment of the full metabeam. {\bf (c)}~Each unit cell has an electronic loop. The voltage $V_s$ induced by the central piezoelectric is fed into a transfer function $H(\omega)= V_{a}(\omega)/V_{s}(\omega)$ that sends opposing voltages $V_a$ and $-V_a$ to the piezoelectric actuators.  {\bf(d)}~A photograph of the metabeam (horizontal) with the electronic circuits in the foreground. We note that the mechanical forces from the attached wires are negligible. The wires act only as sources of energy and computation, but not of linear or angular momentum.
{\bf (e)}~The motion of the metabeam can be described by two independent fields, $\varphi$ and $h$, which parameterize the angular and vertical displacements of the metabeam. Notice  that under a reflection about the $\hat z$ axis, we have $\varphi \to - \varphi$ and $h \to h$. {\bf (f)} When the beam bends, the center piezoelectric is stretched. {\bf (g) } The antisymmetric electronic actuation then gives rise to a shearing stress proportional to the modulus $P$.
}
\label{fig:experiment}
\end{figure*}

\begin{figure}
    \centering
    \includegraphics[width=0.45\textwidth]{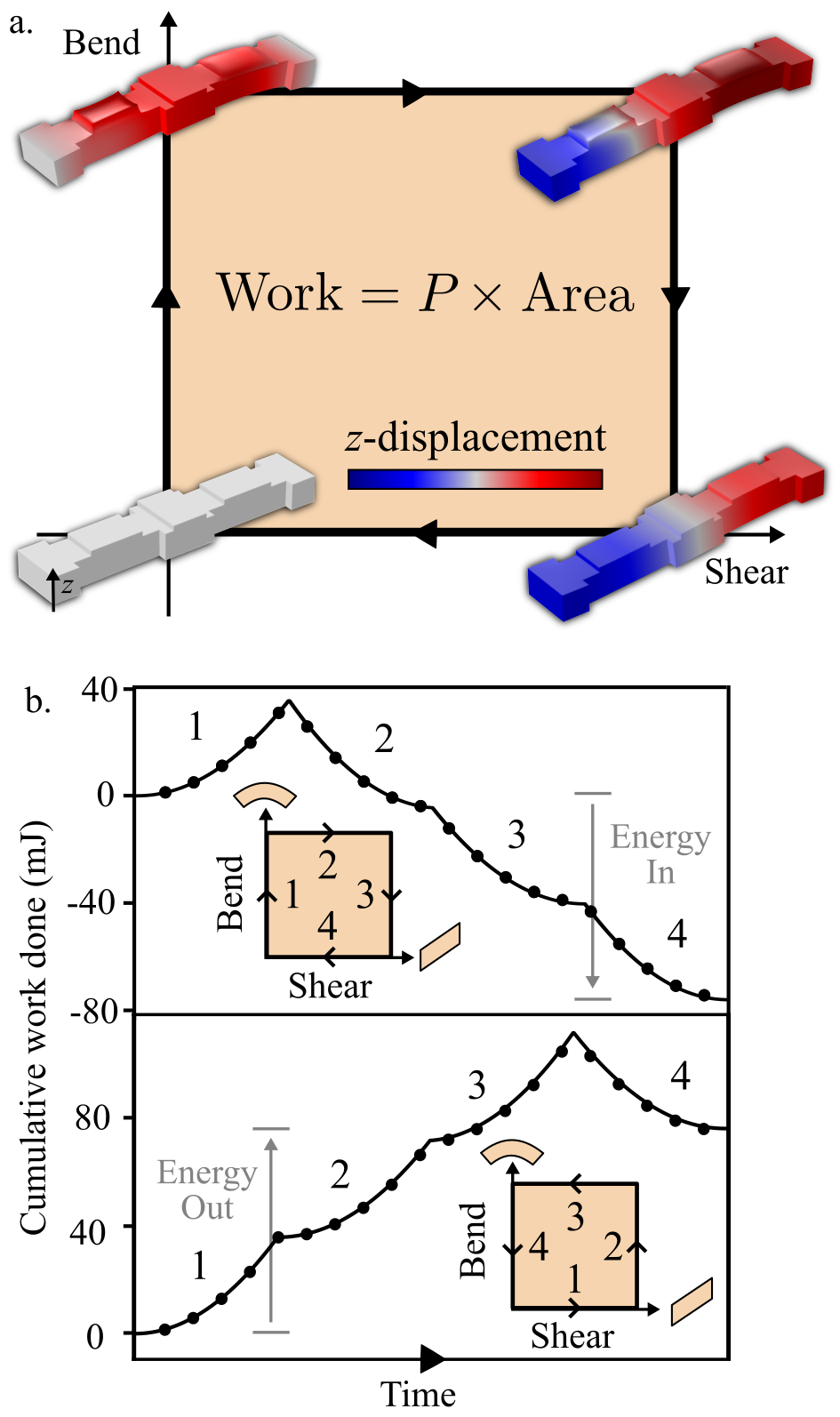}
     \caption{ {\bf Quasistatic deformation cycles with odd micropolar elasticity.}
      {\bf (a)} The state of the unit cell is tracked in the space of shear and bend. When a quasistatic closed path is traced out in this space, the unit cell performs work per unit volume that is proportional to the area times the modulus $P$. {\bf (b)} We numerically compute the work done for a clockwise (top) and a counterclockwise (bottom) path. The solid lines are predictions from the continuum theory, and the black dots result from finite-element simulations of the unit cell. In the simulations, maximum amplitudes of bending and shearing are $b_\text{max} \approx 10^{-1} \si{m^{-1}}$ and $s_\text{max} \approx 10^{-2}$, respectively.  
      See S.I. Section S1H for further details on the simulation. 
     }
    \label{fig:cycle}
\end{figure}

\section{Introduction}

Responsive materials in both biology and engineering distinguish themselves by their ability to respond to external stimuli in tailored ways~\cite{Bertoldi2017,Huber2016,Frenzel2017,Needleman2017,Coulais2016,Guseinov2020,Brandenbourger2019}. For example, muscles contract in response to electrical signals~\cite{caruel2018} and mechanocaloric solids undergo dramatic deformations in response to temperature changes~\cite{Wang2019}. 
Unlike computers or multicellular organisms with specialized functional components, the undifferentiated physical machinery available in a distributed material presents unique challenges for sensing, information processing and response. 
Yet, the range of available functionalities is fundamentally extended when the materials posses distributed, local reservoirs of energy~\cite{Chen2016,Fleury2015, Peng2014,  Li2019particle, Miskin2020, Lin2011,Yi2017,Estakhri2019, Needleman2017,Salbreux2017,prost2015}. Such active materials exhibit responses not allowed by their passive, or energy conserving, counterparts. 

Activity is intimately connected, but not equivalent, to a family of material symmetries collectively known as reciprocity~\cite{Nassar2020, Coulais2017, Fleury2014,fruchart2020phase, saha2020scalar,You2020,gupta2020active, Scheibner2020, Faust2015Reciprocity}.
Here we distinguish between two notions of reciprocity relevant for the design of mechanical metamaterials. 
The first notion is a generalization of Newton's third law, which states that the forces between any two components of a mechanically isolated system must be equal and opposite. This notion of reciprocity can be formulated for other generalized momenta, such as angular momentum. Mechanical systems that violate this version of reciprocity must fundamentally be in contact with an external medium, such as a substrate or a background fluid, to act as a momentum sink. For systems described by a Lagrangian or Hamiltonian, translational (or rotational) symmetry gives rise to conservation of linear (or angular) momentum. Hence, systems with translational symmetry that violate this first notion of reciprocity must either be dissipative, driven, or active.

A second conceptually distinct notion of reciprocity is known as Maxwell-Betti reciprocity, which can be roughly defined as the symmetry between perturbation and response. When a mechanical system is deformed, the work done can schematically be written as $\dd W = \sum_a \sigma_a  \dd u_a$. Here, $u_a$ is a short hand notation for mechanical degrees of freedom such as displacements, rotations or strains, and $\sigma_a$ labels the conjugate forces, moments or stresses. For small perturbations about an undeformed state, we may write $\sigma_a = M_{a b} u_b$. A generic medium is said to obey Maxwell-Betti reciprocity if and only if $M_{a b}$ is symmetric, i.e. $M_{a b} = M_{b a}$. As long as the system's linear response obeys Maxwell-Betti reciprocity, the forces can be derived from gradients of an energetic potential $V = \frac12 M_{ab} u_a u_b$. However, if the linear response of the medium violates Maxwell-Betti reciprocity,  the internal energy is no longer a function of the coordinates $u_a$. In other words, the medium can perform nonzero work along a closed cycle of deformations~\cite{Scheibner2020}. Such a medium necessarily contains non-conservative forces that require an internal or external source of energy to be present. 

While Maxwell-Betti reciprocity and momentum conservation are distinct, one can certainly design and build mechanical systems that harness linear or angular momentum sources to achieve violations of Maxwell-Betti reciprocity~\cite{Brandenbourger2019,Rosa2020,Nassar2020}. Such approaches, however, inherently involve mechanical coupling to an external medium. Here we take an alternative route.  
We explore the case where $\sigma_a$ represents shear and moment stresses, which conserve currents of linear and angular momentum. In this case $u^a$ represents geometric deformations and $M_{ab}$ represents the stiffness matrix containing all of the material's elastic moduli. We refer to the antisymmetric components $(M_{ab}-M_{ba})/2$ as \emph{odd elasticity} \cite{Scheibner2020, scheibner2020non, Zhou2020, banerjee2020active}. A material displaying odd elasticity must violate Maxwell-Betti reciprocity, though it needs not rely on external sources of linear or angular momentum.

Recent advances in metamaterial design and prototyping have utilized active components to achieve functionalities such as sensing, lasing and cloaking~{\cite{Chen2016, Fleury2015,Peng2014,Lin2011}}, frequency dependent reflectivity~\cite{Trainiti2019}, unidirectional wave amplification~\cite{Brandenbourger2019, Fink2000,Sirota2021real}, energy harvesting~\cite{Yi2017} and analog computation~\cite{Estakhri2019}. Nonetheless, all the active non-reciprocal metamaterials so-far realized exhibit either of the following fundamental limitations: the active non-reciprocal effects either vanish from the linear response in the quasistatic limit~\cite{Trainiti2019} or they require the presence of background sources of linear or angular momentum~\cite{Brandenbourger2019,Rosa2020,Sirota2020non}. As a result, their functionalities are largely restricted to finite-frequency control or fundamentally require the sample to be in contact with an additional medium that acts as a momentum sink or source.

Here, we report the design, construction, and experimental demonstration of a freestanding metamaterial whose elasticity is unattainable in passive media (Figure~\ref{fig:experiment}a-d). The metamaterial is constructed with piezoelectric elements~\cite{Ouisse2016piezo,yi2020programmable,
Chen2014jva,Bergamini2014,Alan2019,Chen2017,Wang2017,Casadei2012,Chen2016iop,Sugino2018} mounted on a beam and controlled by electrical circuits. Our approach enables an asymmetric coupling between bending and shearing in micropolar solids~\cite{Eringen1999,Maugin1998}. This results in an odd micropolar material that simultaneously breaks parity and Maxwell-Betti reciprocity. The spectrum of the metabeam exhibits a non-Hermitian topological index which results in the localization of vibrational modes at sample boundaries. 
We experimentally show the resulting unidirectional amplification/attenuation of waves propagating through the metamaterial. Our work sheds light on controlling non-reciprocal elasticity in artificial materials.

\section{Results}

\subsection{Design of the active metamaterial}
The metamaterial we design takes the form of a thick beam, whose shape is characterized by two independent degrees of freedom (Figure~\ref{fig:experiment}e): the height $h(x)$ of the midplane and the angle $\varphi(x)$ of the cross section with respect to the vertical. A single unit cell in the beam is equipped with three piezoelectric patches that enable shape-sensing and response. The central patch acts as a sensor which acquires a voltage proportional to the elongation or contraction of the top surface (Figure~\ref{fig:experiment}f). The piezoelectric patches at the front and back of the unit cell serve as mechanical actuators that elongate and contract in response to an applied voltage~\cite{Chen2018,Chen2020active}. A transfer function $H(\omega)$ processes the input voltage from the central patch and sends output signals to the actuating patches (Figure~\ref{fig:experiment}c). For each unit cell, we implement $H(\omega)$ using a minimal electrical circuit (Figure~\ref{fig:experiment}d). The electronics only couple piezoelectric patches within a single unit cell, thereby creating a control system that is both local and decentralized. 
The resulting system only needs to be connected to a voltage or, more generally, energy source. One practical advantage of this approach is that the voltage or power source can easily be housed inside the medium itself.

The active metamaterial we design is freestanding\textemdash it does not push or pull on an external medium. 
Indeed, it obeys Newton's third law by preserving both angular and linear momentum as a traditional beam. However, the crucial difference between a traditional beam and the one we construct is the presence of internal energy sources. We design the feedback such that when the central patch experiences elongation or compression due to bending of the beam, the electronic loop produces output voltages that are \emph{antisymmetric} (Figure~\ref{fig:experiment}c), resulting in shear stresses (Figure~\ref{fig:experiment}g). However, the ensuring shear strain does not stretch or compress the central piezoelectric patch. Therefore the electromechanical control loop is entirely feed-forward: bending induces shear, while shear does not induce bending.

\subsection{Odd micropolar elasticity} \label{sec:odd}

The feedback results in an elastic response that cannot be realized without an internal source of energy. This effect, apparent in the emergent continuum equations, may be deduced solely using symmetries and conservation laws based on classical beam theory (1D micropolar elasticity). Crucial to our design is the notion of a parity inversion $\PP$, defined here to be a mirror reflection of the beam about the $\hat z$ axis that sends $x$ to $-x$. Figure~\ref{fig:experiment}e shows that under parity, the two independent degrees of freedom, $h(x)$ and $\varphi(x)$, transform as
\begin{align}
h(x) \xrightarrow{\PP}& h(-x) \label{eq:par1} \\
\varphi(x) \xrightarrow{\PP} & - \varphi(-x) \label{eq:par2}
\end{align}
Since $\varphi(x)$ acquires a minus sign under parity, we say that $\varphi(x)$ is a micropolar degree of freedom~\cite{Eringen1999,Maugin1998}. 

The equations of motion for a freestanding micropolar beam are then built out of conservation laws. 
Linear momentum conservation implies that 
\begin{align}
\rho \ddot h = \partial_x \sigma_{zx}, \label{eq:hdd} 
\end{align}
where $\sigma_{zx}$ is the shear stress and $\rho$ is the mass density. Moreover, angular momentum conservation implies
\begin{align}
I\ddot \varphi = \partial_x M + \sigma_{zx}, \label{eq:phidd}
\end{align}
where $M$ is the bending moment and $I$ is the cross sectional moment of inertia. The moment 
$M$ and stress $\sigma_{zx}$ are themselves determined by the deformation of the beam via a set of constitutive relations. To leading order in gradients of $h$ and $\varphi$, the internal geometry of the beam is approximated by two independent types of deformation: bending $b(x)$ (Figure~\ref{fig:experiment}f) and shearing $s(x)$ (Figure~\ref{fig:experiment}g), 
defined as
\begin{align}
    b(x) = & \partial_x \varphi \\
    s(x) = & \partial_x h - \varphi 
\end{align}
Under parity inversions, Eqs.~(\ref{eq:par1}-\ref{eq:par2}) imply that
\begin{align}
    b(x) &\xrightarrow{\PP} b(-x) \\
    s(x) &\xrightarrow{\PP} -s(-x) 
\end{align}

Assuming a linear response, one may in general write the linear constitutive relations as:
\begin{align}
    \mqty[ \sigma_{zx} (t)  \\ M(t) ] = \int_{-\infty}^\infty \mqty[ C_{11} (t') & C_{12} (t') \\ C_{21} (t') & C_{22}(t') ]  \mqty[ s(t-t') \\  b(t-t')  ] \dd t' \label{eq:mattime}
\end{align}
or in terms of frequency:
\begin{align}
    \mqty[  \sigma_{zx} (\omega) \\ M (\omega) ] 
    = \mqty[ C_{11} (\omega) & C_{12} (\omega) \\  C_{21} (\omega) & C_{22} (\omega) ] 
    \mqty[ s(\omega) \\ b(\omega)  ].
    \label{eq:mat}
\end{align}
We denote the matrix operator on the right-hand side of Eq.~(\ref{eq:mattime}) as $\mathbf{C}(t)$. Causality implies that $\mathbf{C}(t) =0$ for $t<0$ and reality of $\mathbf{C}(t)$ implies $\mathbf{C}(-\omega) = \mathbf{C}^*(\omega)$. Direct substitution of Eq.~(\ref{eq:mat}) into Eqs.~(\ref{eq:hdd}-\ref{eq:phidd}) reveals that the equations of motion are invariant under parity if and only if $C_{21}$ and $C_{12}$ are zero. Hence we say that $C_{21}$ and $C_{12}$ are micropolar moduli. 

It is useful to parameterize the $\mathbf{C}(\omega)$ as
\begin{align}
    \mqty[ C_{11} (\omega) & C_{12} (\omega) \\  C_{21} (\omega) & C_{22} (\omega) ] = \mqty[ \mu(\omega) & \alpha(\omega) + \beta(\omega) \\ \alpha(\omega) - \beta(\omega) & B(\omega) ] \label{eq:param} 
\end{align}
where $\mu$ is the shear modulus and $B$ is the bending modulus, and $\alpha$ and $\beta$ are the symmetric and antisymmetric components of the micropolar moduli. If the beam lacks an internal source of energy, then the total work done by the beam on any deformation process that begins and ends in the same configuration must be less than zero: $\Delta W \le 0$. 
This energy condition places the following constraints on the finite frequency linear response coefficients (see Methods):
\begin{align}
   0 \ge & \Im[\mu(\omega) +B(\omega) ] \label{eq:cond1}  \\
   0 \le & \Im[\mu(\omega) ]\Im[B(\omega)] - \Im[\alpha (\omega) ]^2 -\Re[ \beta(\omega)]^2\label{eq:cond2} 
\end{align}
for all $\omega$. 
Notice that the reality condition $\mathbf{C}(-\omega) = \mathbf{C}^*(\omega)$ implies $\Im[\mathbf{C}]=0$ at $\omega=0$. As a consequence, we must have $\beta(\omega \to 0) =0$  
for any passive beams.

However, our active metamaterial has an internal source of energy and thus need not obey this constraint. The feed-forward coupling between bending and shearing suggests a linear response matrix of the form:
\begin{align}
    \mqty[ C_{11}(\omega) & C_{12} (\omega)  \\  C_{21} (\omega) & C_{22} (\omega) ] = \mqty[ \mu (\omega) & P (\omega) \\ 0 & B (\omega) ] \label{eq:form}
\end{align}
The electronic control loop introduces the coefficient $P = 2\alpha = 2\beta$, which we refer to as the odd micropolar modulus. This modulus breaks two crucial symmetries. Since the electromechanical coupling violates parity, the active modulus $P$ must occur in the off-diagonal entries. Moreover, $P$ occurs only in the upper-right entry because the electro-mechanical coupling is feed-forward: bend causes shear, but shear does not cause bend. As a result the matrix $\mathbf{C}$ is asymmetric, indicating that the beam violates Maxwell-Betti reciprocity, even at zero frequency.

For our metabeam, we measure the moduli via COMSOL simulations in which we apply controlled displacements at finite frequency to the front and back faces of a single unit cell. By measuring the reaction forces on these faces, we determine the resulting stresses and, consequently, the moduli. We empirically find that $\mu = 1.3 \times 10^9 \si{kg / m s^2}$ and $B=0.112 \times 10^6 \si{kg/s^2}$ are approximately independent of frequency, while $P(\omega) = \Pi H(\omega)$, where $H(\omega)$ is the transfer function and $\Pi = 4.7 \times 10^6 \si{kg m/ s^2} $ is a material constant. 
From the metabeam geometry and materials, we compute the average volumetric density $\rho = 5,613 \si{kg /m^3} $ and the average cross-sectional moment of inertia $5.9 \times 10^{-3} \si{kg/m}$. See the Supplementary Information (S.I.) for additional characterization details.

\begin{figure}
    \centering
    \includegraphics[width=\columnwidth]{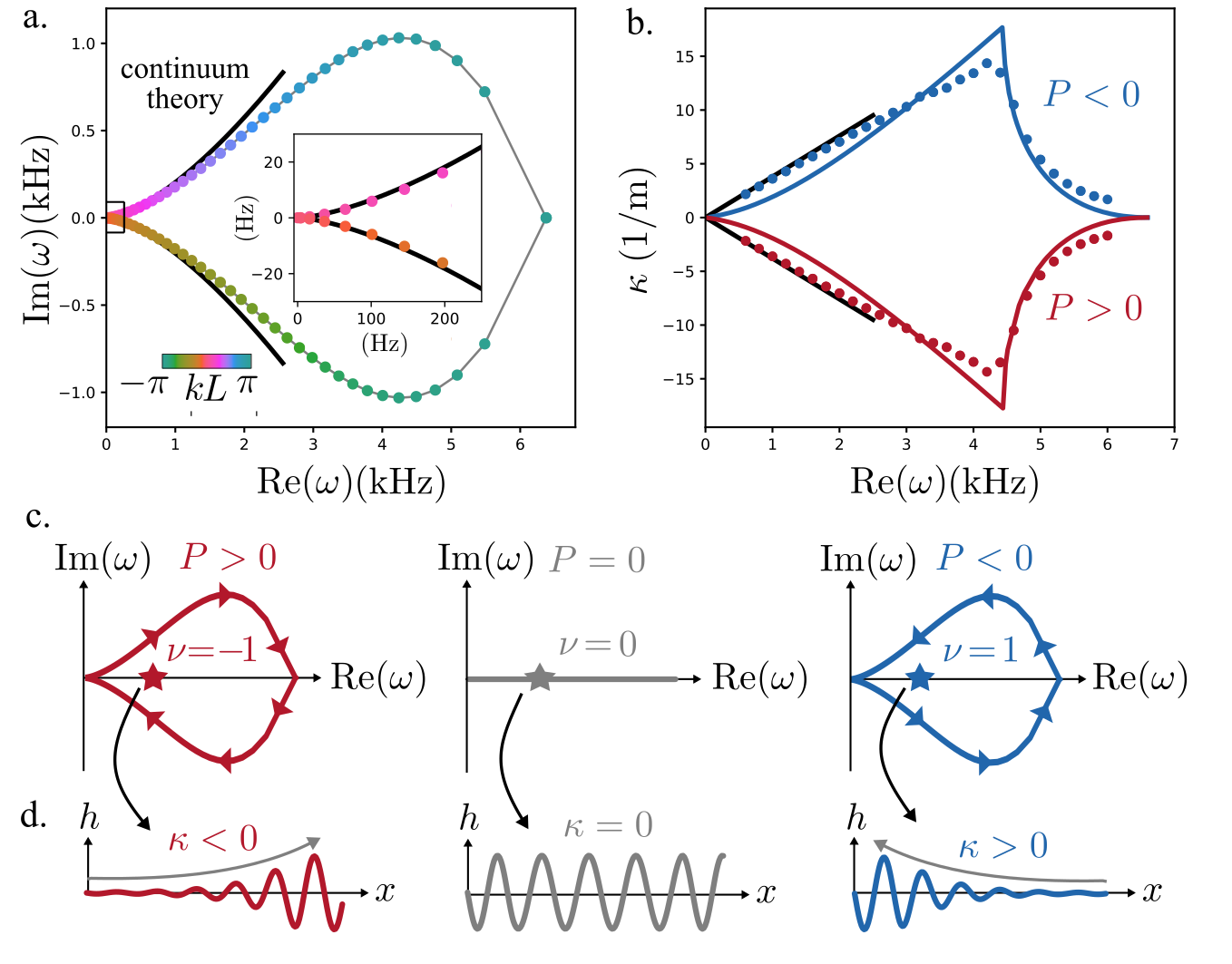}
    \caption{{\bf Non-Hermitian skin effect via the odd micropolar elasticity.}~{\bf (a)}~The vibrational spectrum for the flexural mode of a metabeam with periodic boundary conditions and $P = 3 \Pi$. The black line results from the continuum theory given by Eq.~(\ref{eq:approx1}) with parameters given in Sec.~\ref{sec:odd}. The data points are obtained via fully piezoelectrically coupled simulations in COMSOL with the hue indicating the wavenumber $k$. 
    For the full spectrum plotted as a function of $k$ in the continuum theory and in the numerics, see Figures~\ref{fig:contwind} and S2 respectively. The inset compares the continuum theory and simulations for small wavenumbers $k \ll 1/{\sqrt{\ell_1 \ell_2}}$~{\bf (b)}~The penetration depth for real $\omega$ in a medium with open boundary conditions.  The points are the results of COMSOL simulations, the black lines are Eq.~(\ref{eq:kappa}), and the dark lines are the result of the transfer matrix method, see S.I. Section S3A~{\bf (c)}~The localized states are connected to a topological index $\nu(\omega)$. The periodic boundary spectrum for $P>0$, $P=0$, and $P<0$ are represented schematically by the solid lines. The arrows indicate the direction of increasing $k$. For a given frequency $\omega$, the winding number $\nu(\omega)$ of the periodic boundary spectrum indicates the presence of a localized mode. {\bf(d)}~The localization of eigenmode at the value of $\omega$ denoted by the star in panel (c) is schematically illustrated.
    }
    \label{fig:numer_main}
\end{figure}

\subsection{Energy cycles} 
It is useful to consider the elastic limit of Eq.~(\ref{eq:param}) in which $\mathbf{C}(\omega)$ is real and approximately independent of frequency. For a passive beam in this limit, the matrix $\mathbf{C}$ is obtained by approximating the energetic cost of deformation by a quadratic function:  
\begin{align}
    W= \frac{\mu}2 s^2 + \frac{B}2 b^2 + \alpha s b \label{eq:alpha}
\end{align}
The work (per unit volume) done in an infinitesimal deformation of the beam is given by:
\begin{align}
    \dd W = \sigma_{zx} \dd s + M \dd b \label{eq:work}
\end{align}
From Eq.~(\ref{eq:work}), we conclude $\sigma_{zx} = \pdv{W}{s}$ and $M= \pdv{W}{b}$. Hence we obtain the linear constitutive relations:
\begin{align}
\mqty[ \sigma_{zx} \\ M] = \mqty[ \mu & \alpha \\ \alpha & B] \mqty[ s \\ b] \label{eq:const1}   
\end{align}
Since $\mathbf{C}$ in Eq.~(\ref{eq:const1}) is obtained via a second derivative of an energy function, it is symmetric. This is a manifestation of the Maxwell-Betti reciprocity theorem. In this case, the cumulative work done over a sequence of deformations depends only on the initial and final configurations: $\Delta W = \int \dd W = W_\text{final} - W_\text{initial}$. In particular, if the procedure begins and ends at the same state, we have $\Delta W=0$.

Notice that the constitutive relation in Eq.~(\ref{eq:form}) violates $\mathbf{C} = \mathbf{C}^T$. Hence the Maxwell-Betti reciprocity theorem implies that the form of $\mathbf{C}$ in Eq.~(\ref{eq:form}) does not follow from a potential energy function. To see this explicitly, consider the differential of energy given by:
\begin{align}
    \dd W =&  \sigma_{zx} \dd s +  M \dd b \\
    =& \dd( \frac{\mu}2 s^2 + \frac{B}2 b^2 + \frac{P}2 s b ) + \frac{P}2 (b \dd s -s \dd b).  \label{eq:diff2}
\end{align}
Notice that the second term in Eq.~(\ref{eq:diff2}) cannot be expressed as the differential of a potential. By Green's theorem, integrating the work done $\oint \dd W $ over a closed loop in strain space yields: 
\begin{align}
\oint_{\partial_V} \dd W =& \oint_V  P \dd s \dd b =P \times \text{Area}
\end{align} 
where ``$\text{Area}$" is the signed area of the region $V$ enclosed by the path in the space of strains.  
Notice that the work done per cycle can be either positive or negative depending on the orientation of the path, and is independent of the rate of the process (in the quasistatic limit). 
Since the crucial ingredient for such cycles is an antisymmetry in $\mathbf{C}$, we refer to this form of elasticity as ``odd" (i.e. antisymmetric) elasticity~\cite{Scheibner2020}.

To verify that our design displays this property at low frequencies, we perform COMSOL simulations of the beam with full piezoelectric coupling. As illustrated in Figure~\ref{fig:cycle}a, we subject a single unit cell to a four-step protocol of shearing and bending by enforcing displacement controlled boundary conditions at the two ends of the beam. We measure the reaction forces on the control surfaces to compute the work done by the beam, plotted in Figure~\ref{fig:cycle}b. 
When the deformations are performed in a clockwise direction in strain space, as shown in the top panel of Figure~\ref{fig:cycle}b,  the cumulative work done is negative once the unit cell returns to its initial configuration. Hence, energy flows from the external agent into the internal power reserves of the medium. When the cycle is reversed, so is the flow of energy. The ability to inject or extract mechanical energy through quasistatic cycles is synonymous with odd elasticity, i.e. a quasistatic stress-strain relationship that does not follow from a potential energy.
In the quasistatic limit, the total work done is given by $P$ times the area enclosed in strain space. 
In the Methods, we also derive energy relations for cycles at finite frequency. For those cases, the total work done is related to $\abs{P}$, $\arg(P)$, and the area enclosed in strain space.

\begin{figure}
    \centering
    \includegraphics[width=0.48\textwidth]{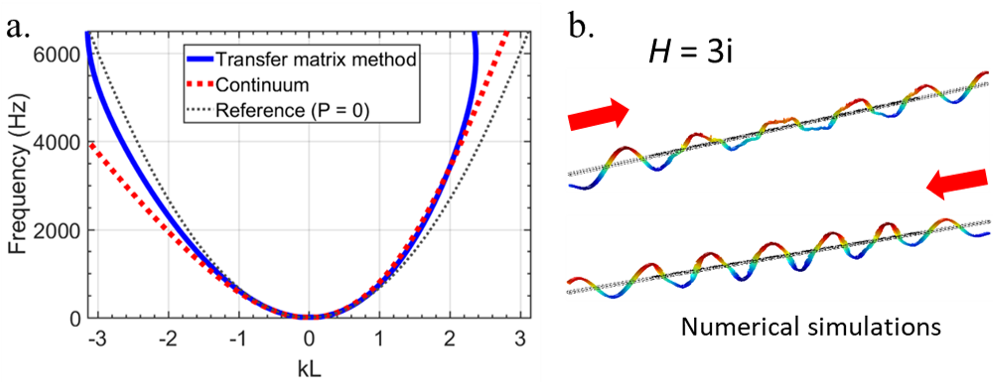}
    \caption{{\bf Pseudo-Hermitian dynamics.}~{\bf (a)}~ The spectrum is shown for the metamaterial with $\arg(P) = \pi/2$. We note that the reality of the frequencies is maintained, while the modulus $P$ breaks the $k \mapsto - k$ symmetry. {\bf (b)}~ Transverse displacement wave fields for the waves traveling in different directions. The left and right traveling modes are excited at equal frequencies, but have differing wavenumbers due to the odd micropolarity.  
    }
    \label{fig:Hermitian}
\end{figure}

\subsection{Odd micropolar elastodynamics and the non-Hermitian skin effect} 

We now ask about the dynamic consequences of the active feed-forward control with parity violation. We first consider the elastic approximation in which $\mathbf{C}(\omega)$ is real and frequency independent. The linearized continuum equations governing the motion of the beam are given by:
\begin{align}
    \rho  \ddot h 
    =& \mu \partial_x^2 h + P \partial_x^2 \varphi - \mu \partial_x \varphi \label{eq:dyn1}\\
    I \ddot \varphi 
    =& \mu \partial_x h + B \partial_x^2 \varphi + P \partial_x \varphi - \mu \varphi\label{eq:dyn2}
\end{align}
Using Fourier transform, Eqs.~(\ref{eq:dyn1}-\ref{eq:dyn2}) may be cast in the form:
\begin{align}
{\omega}
{\begin{bmatrix}
\tilde p_h \\ \tilde p_\varphi \\ \tilde s \\ \tilde b 
\end{bmatrix}}{=} 
\underbrace{
{\omega_1}{\begin{bmatrix}
 0 & 0 & - k \ell_1 & - \tilde P \ell_1 k \\
0& 0 & i & i \tilde P  - k  \ell_2 \\
-k \ell_1 & -i & 0 & 0 \\
0 & -k \ell_2 & 0 & 0
\end{bmatrix}} }_{\dD (k)}
{\begin{bmatrix}
\tilde p_h \\ \tilde p_\varphi \\ \tilde s \\ \tilde b 
\end{bmatrix}} 
\label{eq:Ddef}
\end{align}
where $k$ and $\omega$ are the wave number and frequency associated with the Bloch wave $e^{i (kx-\omega t)}$. We have introduced the notation: 
\begin{align}
    \tilde s =& \sqrt{\mu}(\partial_x h - \varphi) 
     & \tilde b =& \sqrt{B} \partial_x \varphi \\
    \tilde p_\varphi =& \sqrt{I} \dot \varphi & 
    \tilde p_h =& \sqrt{\rho} \dot h  \label{eq:basis}
\end{align}
Here, $\tilde s$ represents the shear, $\tilde b$ represents the bending, $\tilde p_\varphi$ is the angular momentum, and  $\tilde p_h$ is the $z$-component of the linear momentum. This parameterization is natural since the standard inner product 
\begin{align} 
2e = \abs{\tilde p_h}^2 + \abs{\tilde p_\varphi}^2+ \abs{ \tilde s}^2 + \abs{\tilde b}^2  \label{eq:inner} 
\end{align}
is equal to twice the mechanical energy density $e$. The dynamical matrix $\dD(k)$ depends on four parameters: $\ell_1 \equiv \sqrt{I/\rho} $ is roughly the thickness of the metabeam; $\ell_2 \equiv \sqrt{B/\mu}$ is the distance over which shearing and bending of equal transverse deflection cost equal amounts of energy;  $\omega_1 \equiv \sqrt{\mu/I}$ sets a frequency scale separating transverse flexural modes and high frequency shearing modes (See Figure~\ref{fig:contwind}a); finally, the parameter $\tilde P \equiv P/\sqrt{B \mu} $ is the normalized odd micropolar modulus. For our metabeam $\ell_1 \approx 10^{-3} \si{m} $, $\ell_2 \approx 10^{-2} m$, $\omega_1 \approx 10^5 \si{Hz}$, and $\abs{\tilde P} \lesssim 1$. 

Within the continuum theory, the vibrational dynamics can be captured by solving the secular equation $\det[\dD(k) -  \omega] =0$, which takes the form
\begin{align}
    0= \tilde \omega^4 -  \qty[ 1- i \tilde P k \ell_2 +k^2 (\ell_1^2 + \ell_2^2) ] \tilde \omega^2 + k^4 \ell_1^2 \ell_2^2 \label{eq:sec}
\end{align}
where $\tilde \omega = \omega /\omega_1$. Notice that Eqs.~(\ref{eq:dyn1}-\ref{eq:dyn2}) are two coupled second order equations and hence permit a dispersion with four branches. For small wavenumber, the dispersion for the low frequency flexural bands is given by:
\begin{align}
    \omega_\pm = \pm \omega_1 \qty[ \ell_1 \ell_2 k^2 \pm i \tilde P  \ell_1 \ell_2^2 k^3 + \order{ \ell_1^2 \ell_2^2 k^4} ] \label{eq:approx1} 
\end{align}
As can be seen from Eq.~(\ref{eq:approx1}), when $P$ is nonzero and real, the periodic boundary spectrum acquires a nonzero imaginary component. The nonzero imaginary contribution arises since the active metabeam has the ability to physically introduce or remove mechanical energy. 
Moreover, the modulus $P$ violates parity and hence breaks the symmetry between $k \to - k$, in contrast to the design in Ref.~\cite{Brandenbourger2019}, where the parity violation is induced by a term in the dispersion that is linear in $k$. 

The simultaneous breaking of parity and energy conservation allows our active micropolar metamaterial to selectively amplify and attenuate waves based on their direction of travel. 
In Figure~\ref{fig:numer_main}a, we show the spectrum of the flexural mode in the right half of the complex plane for $P>0$. The solid line is the continuum theory, valid at small $k \ll 1/\sqrt{\ell_1 \ell_2}$, and the discrete points are the results of fully coupled COMSOL simulations. In the calculations, $P = \pm 3 \Pi$, and there are no free fitting parameters. (The material constants $\Pi$, $\rho$, $B$, and $\mu$ in the continuum theory are determined from independent simulations, see Section~\ref{sec:odd}.)
As illustrated in Figure~\ref{fig:numer_main}a, for $P>0$, $\Im(\omega ) >0$ whenever $\Re(\dd\omega /\dd k)>0$ and $\Im(\omega ) <0$ whenever $\Re(\dd\omega /\dd k)<0$. Physically, this means that wave packets traveling to the right are amplified, while wave packets traveling to the left are attenuated. As an illustration, Figure~\ref{fig:numer_main}b shows the inverse penetration depth $\kappa$ for $\omega$ taken to lie along the positive real axis. From the continuum theory, we compute the analytical formula for $\kappa$ (see Methods)
\begin{align}
    \kappa = \sqrt{\frac\rho B} \frac{P}{4 \mu} \omega \label{eq:kappa}
\end{align}
In Figure~\ref{fig:numer_main}b, we compare Eq.~(\ref{eq:kappa}) to the COMSOL simulations (see Methods) and a semi-analytical technique known as the transfer matrix method (see S.I. Section S3A). 

This unidirectional amplification can be understood from the point of view of the non-Hermitian skin effect~\cite{Hatano1996,Bergholtz2021Exceptional,ashida2020nonhermitian,trefethen2020spectra, kawabata2020topological,ghatak2019,Lee2019,Helbig2020,Yao2018,Okuma2020,Hofmann2020Reciprocal, Zhang2020Correspondence}. Given a complex frequency $\omega$, we can define the following topological index:
\begin{align}
    \nu(\omega) = \frac{1}{2 \pi i} \sum_{\alpha }  \oint_{- \pi /L}^{\pi /L} \dv{k} \log[\omega_\alpha(k) - \omega ] \dd k \label{eq:wind1}  
\end{align}
where $L$ is the length of a single unit cell and $\omega_\alpha (k)$ is the frequency of the $\alpha$ band. Here, we take $\alpha $ to run over the flexural bands. 
The topological index $\nu(\omega)$ indicates whether a system with semi-infinite boundary conditions will host a localized mode at the frequency $\omega$. When $\nu(\omega) >0$, a semi-infinite system with domain $x \in [ 0, \infty)$ will host a mode localized to its left boundary at frequency $\omega$. Likewise, when $\nu(\omega) < 0$, a semi-infinite system with domain $x \in (- \infty, 0]$, will host a mode localized to its right boundary.

As shown in Figure~\ref{fig:numer_main}c-d, the sign of $\kappa$ can be rationalized by examining $\nu(\omega)$ for an example frequency $\omega$ (denoted by the star) along the real axis. For $P>0$, the periodic boundary spectrum (red) winds once  counterclockwise around the star, and hence $\nu(\omega) < -1$. However, for $P<0$, the localization is reversed since the direction of the contour is reversed.  In the Methods,  we show how to compute $\nu(\omega)$ directly from the continuum equations using a generalization of Eq.~(\ref{eq:wind1}). Furthermore, in S.I. Section S1G we discuss how the presence of additional vibrational bands affect the calculation and physical interpretation of $\nu(\omega)$.

\subsection{Pseudo-Hermitian dynamics and direction-dependent bending modulus} 

In the preceding section, we took $P(\omega)$ to be real and constant. However, by introducing a phase delay into the transfer function $H(\omega)$ we can control the complex argument of $P$. When $\arg(P) = \pm \pi/2$, the secular Eq.~(\ref{eq:sec}) has entirely real coefficients. Hence, the periodic boundary spectrum will consist of frequencies that come in real values or complex conjugate pairs. This additional symmetry is sometimes referred to as a generalized PT symmetry~\cite{Mostafazedeh2015Spectral,Bender1998,scheibner2020non}, which arises if and only if there exists an antiunitary operator that commutes with $\dD(k)$. We say that the PT symmetry is \emph{unbroken} when the eigenvalues of $\dD(k)$ are entirely real, and that it is broken otherwise. In the unbroken phase,  $\dD(k)$ is said to be \emph{pseudo}-Hermitian.  Pseudo-Hermiticity implies that each eigenvector of $\dD(k)$ will individually conserve the mechanical energy density $e$ in Eq.~(\ref{eq:inner}). However, when two or more eigenvectors are superimposed, $e$ can oscillate in time, though remaining centered around a constant time-averaged value. 

Since pseudo-Hermiticity constrains the periodic boundary spectrum of $\dD(k)$ to lie along the real line, the unidirectional amplification vanishes when $P$ is imaginary. 
Nonetheless, the effects of parity violation are still present. We note that Eq.~(\ref{eq:approx1}) may be written in the form:
\begin{align}
    \omega_\pm  = \pm \sqrt{\frac{B}{\rho}} \qty( 1 \pm i \frac{Pk}{\mu})k^2 + \order{ k^4 \frac{I^2 B^2 }{\rho^2 \mu^2} }
\end{align}
When $\arg(P) = \pm \pi/2$, we can interpret the form of Eq.~(\ref{eq:approx1}) as having a rescaled bending modulus: 
\begin{align}
    B_{\text{eff}}= B \qty(1 \mp \frac{\abs{P} k }{2 \mu})^2 \label{eq:B}
\end{align}
Notice that the value of $B_\text{eff}$ depends on the sign of $k$, and hence the effective bending modulus is direction dependent.  
Figure~\ref{fig:Hermitian}a shows the tilting of the dispersion for $\arg(P)=\pm \pi/2 $. The tilt implies that the phase and group velocities for right and left traveling waves are unequal. Numerically solved modes are shown in Figure~\ref{fig:Hermitian}b. We note that the pseudo-Hermiticity endowed by  $P(\omega) \propto \pm i$ must exist exclusively at finite frequency because $P(\omega)$ cannot be nonzero and imaginary at $\omega=0$ due to the requirement that $P(-\omega)=P^*(\omega)$. 

\begin{figure}
    \centering
    \includegraphics{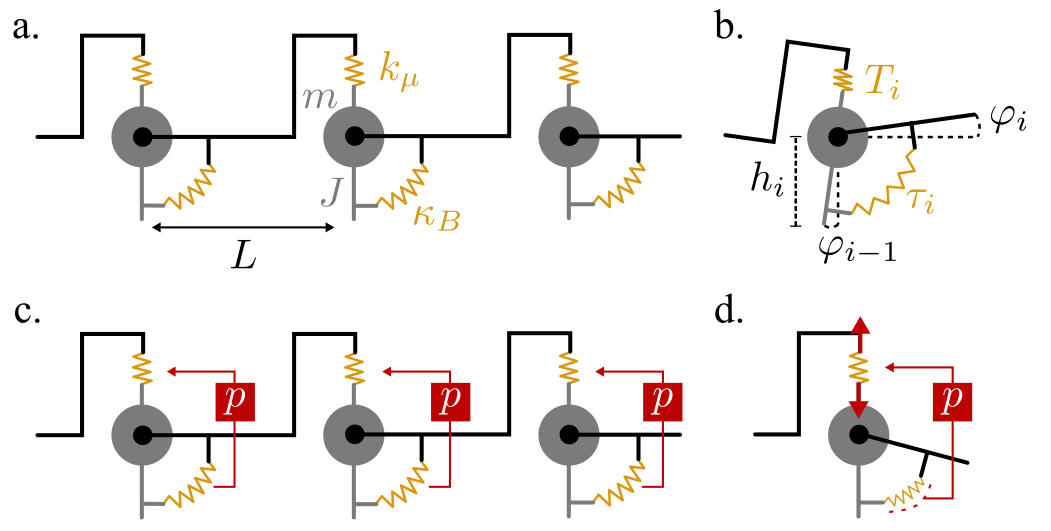}
    \caption{ {\bf Discrete model for odd micropolar beam.}~{\bf (a)} A discrete model of a Timeoshenko beam consists of a central mass $m$ with moment of inertial $J$, a Hookean spring of spring constant $k_\mu$ and a torsional spring of spring constant $\kappa_B$ and lattice spacing $L$.~{\bf (b)} The unit cell is described by the height of the mass $h_i$ and the angle $\varphi_i$ of the black connecting rod.~{\bf (c-d)} The odd micropolar beam has an internal feedback that senses the angle change of the torsional spring and actuates additional tension in the Hookean spring. The control loop is  unidirectional: stretching or compressing the Hookean spring does not affect the torsional spring.}     
    \label{fig:desc}
\end{figure}

\subsection{  A discrete model of the odd micropolar metabeam} 
\label{sec:desc} 
To gain intuition into the mechanics of the metabeam, it is useful to consider a discrete model. As shown in Figure~\ref{fig:desc}a-b, the $i$th unit cell of the discrete model consists of a rod (gray) with moment of inertia $J$ and total mass $m$, whose position and orientation are captured by a height $h_i$ and an angle $\varphi_{i-1}$. The rods are connected via massless, rigid frames and two springs. The top spring is a Hookean spring with tension 
\begin{align}
T_i = k_\mu (h_{i-1} - h_{i} + L \varphi_{i-1}  ) 
\end{align} 
and a bottom spring is a torsional spring that exerts an angular tension 
\begin{align}
\tau_i = \kappa_B (\varphi_{i-1} - \varphi_{i}) 
\end{align} 
In Figure~\ref{fig:desc}c-d, we show the addition of an active element that senses the stretching of the bottom spring and actuates an additional tension in the top spring 
\begin{align}
T^a_i = p ( \varphi_{i-1}-\varphi_{i} ) 
\end{align} 
Summing the forces in the vertical direction yields a dynamical equation for $h_i$. 
\begin{align}
    m \ddot h_i =&  T_{i} -T_{i+1} +T^a_{i} - T^a_{i+1} \\
                =&  k_\mu (  h_{i+1} +h_{i-1} - 2 h_i) + L k_\mu (   \varphi_{i-1} -\varphi_{i}  ) \nonumber \\
                 &+ p ( \varphi_{i+1} + \varphi_{i-1} - 2 \varphi_i  ) \label{eq:desc1} 
\end{align}
Furthermore, summing the torques yields
\begin{align}
    J \ddot \varphi_i =&  \tau_i -\tau_{i+1}  - L (T_{i+1} + T^a_{i+1} ) \\
    =& \kappa_B ( \varphi_{i+1} + \varphi_{i-1} - 2 \varphi_i  )  + pL (\varphi_{i+1} - \varphi_i ) \nonumber \\
    &+ L k_\mu ( h_{i+1} - h_{i} - L \varphi_{i} ) \label{eq:desc2}
\end{align} 
Upon inspection, Eqs.~(\ref{eq:desc1}) and (\ref{eq:desc2}) are precise discretizations of Eqs.~(\ref{eq:dyn1}) and (\ref{eq:dyn2}) with $\rho = m/L$, $I= J /L$, $\mu= k_\mu L $, $ B= \kappa_B L$, and $P = L p$. See Fig.~S4 for a comparison of the dispersion for the discrete model and continuum theory. Notice that the discrete model manifestly conserves linear and angular momentum since the linear and torsional springs exert equal and opposite forces and torques, respectively, on the units they connect. Even without externally applied torques, nontrivial internal angular momentum transfer occurs between the translation of its center of mass $ j L m \dot h_j$  and the rotation of the axis $J \dot \varphi_j$, akin to a ``spin-orbit" coupling. Nonetheless, Maxwell-Betti reciprocity is violated by the asymmetry in the relationship between the linear and torsional springs: bending of the torsional spring induces a tension $T^a_i = p (\varphi_{i-1} - \varphi_i ) $ in the linear spring, while the deformation of the linear spring $h_{i-1} -h_i+L \varphi_{i-1}$ has no response in the angular spring. This asymmetry implies that a cycle of alternating actuation and release of the linear and torsional spring, analogous to that shown in Fig.~\ref{fig:contwind}a, is associated with a nonzero amount of work done. 
For additional discrete models illustrating the independence of Maxwell-Betti reciprocity and momentum conservation, see S.I. Section S1B.

\subsection{Experimental demonstration} \label{sec:expt}

 \begin{figure*}
    \centering
    \includegraphics[width=0.75\textwidth]{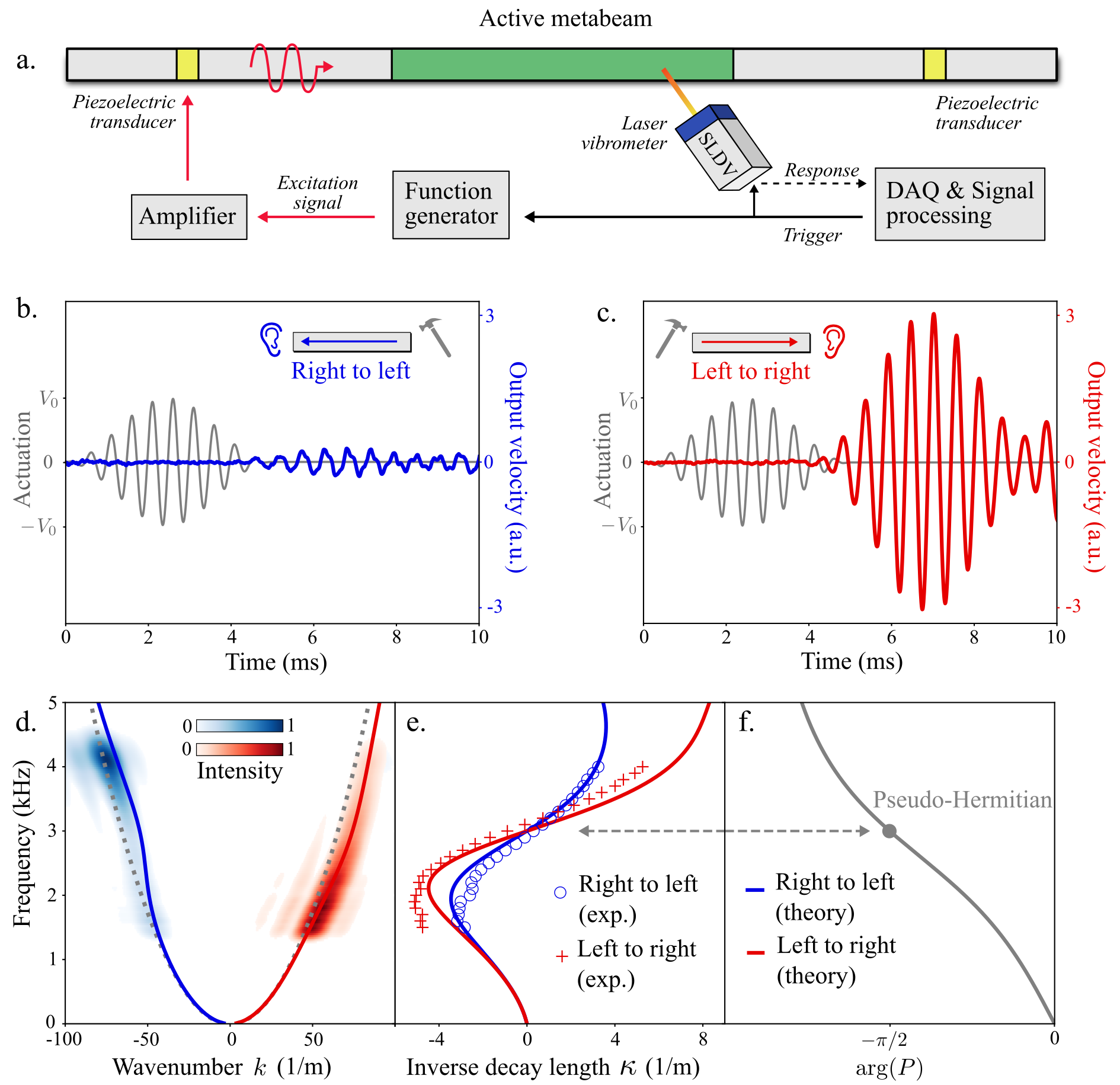}
     \caption{{\bf Experimental demonstration of skin modes and odd micropolar moduli.} {\bf (a)} Experimental schematic. Flexural waves are generated in the active metabeam from either the right or left side using piezoelectric actuators (yellow), see Methods. A scanning laser Doppler vibrometer (SLDV) measures the transverse velocity of the surface of the active metabeam. {\bf (b-c)}~Unidirectional amplification of waves. A metamaterial consisting of 9 unit cells is actuated from either the right (blue) or left (red) with a 2~kHz tone burst signal (grey). The output velocity is measured in units of the peak velocity for experiments with no active feedback.  {\bf (d)}~ Observation of the non-Hermitian skin effect. Experiments are performed between 1.5 kHz and 4 kHz for right to left (blue) and left to right (red) traveling waves. A 2D FFT shows the intensity of the observed spectrum.  {\bf (e)}  The inverse decay length. In (d) and (e), the solid theoretical curves are based on the transfer matrix method. In (d), the grey dashed curves are theoretical predictions with no activity. {\bf (f)}  A plot of $\arg(P)$ as a function of frequency. At $\omega=\omega_0(=3$~kHz$)$, $\arg(P)=-\pi/2$, indicating that the system is pseudo-Hermitian and accordingly we observe $\kappa=0$ at $\omega =\omega_0$.}. 
    \label{fig:Amp}
\end{figure*}

To probe the dynamic wave phenomena originating from the odd micropolar modulus $P$, we perform experiments in which we excite flexural waves in the metabeam using piezoelectric actuators, see Figure~\ref{fig:Amp}a and Methods.   
In experiments, we implement the following electronic transfer function (cf. Figure~\ref{fig:experiment}c): 
\begin{align}
    H(\omega) = \frac{H_0}{(i \omega/\omega_0)^2 +2i \zeta \omega/\omega_0 +1}. \label{eq:h}
\end{align}
Here, $\zeta =0.48$ and $H_0=3$ are constants and $\omega_0=3$~kHz is the cutoff frequency at which $\arg(P) = -\pi/2$.
Other electrical circuit and geometric parameters of the metabeam can be found in S.I. Section S2B.
We probe the vibrational dynamics of the beam by initiating waves from either the right or left side of the metamaterial via external piezoelectric elements. Figure~\ref{fig:Amp}b-c shows the experiment at 2~kHz in which waves from the right are suppressed while waves from the left are amplified (see also Supplementary Movies S1 and S2). 

To construct the full spectrum of the metamaterial, we perform the experiment with tone burst signals centered between $1.5$ and $4$~kHz. 
The transverse velocity wave fields are measured along the medium using a laser Doppler vibrometer. We apply fast-Fourier transforms in time and in space to extract the real $k(\omega)$ and imaginary part $\kappa(\omega)$ of the wavenumber as a function of frequency for right-going (red) and left-going (blue) waves (Figure~\ref{fig:Amp}d-e).  As described in S.I. Section S3A, the solid theoretical curves are produced using a semi-analytical technique known as the transfer-matrix method. The transfer-matrix method utilizes the beam geometry, known material parameters, and electronic feedback measured in simulations. No fitting parameters are used in the comparison between experiment and the transfer-matrix method curve.  

As illustrated in Figure~\ref{fig:Amp}f, the transfer function $H(\omega)$ is chosen such that $-\pi/2 < \arg(P) <0$ when $\omega < \omega_0$. In this case, we find $\kappa <0$ for both left- and right-propagating waves (panel e). A value of $\kappa <0$ implies that waves propagating to the left are attenuated whereas waves propagating to the right are amplified, which is confirmed by the FFT intensity in panel d. Likewise  $-\pi < \arg(P) < -\pi/2$ for $\omega > \omega_0$. In this case, we find $\kappa > 0$, indicating that waves propagating to the right are attenuated whereas waves propagating to the left are amplified. In addition, when $\omega = \omega_0$, $\arg(P) = -\pi/2$, and the waves propagating to the left and right display no attenuation or amplification. At this frequency, the effective dynamical matrix is pseudo-Hermitian point, and the differences between left- and right-propagating waves reside only in the wavelengths and phase velocities. We note that, in experiments, the unidirectional amplification is limited by the maximum output voltage ($\pm 45$V) of our electric control system. In practice, to maximize the amplification ratio, one strategy is to use a modest value of the transfer function, say $\abs{H_0} \approx 3$, and increase the number of unit cells over which the wave is amplified.

\section{Discussion}

The metabeam presented here demonstrates active odd micropolar moduli and non-reciprocal responses absent in energy conserving media that are enabled by sensing, actuating and local computation. The minimal on-board electronics that power the active metabeam enable its multiple functions as an elastic engine, selective mode amplifier, and mechanical diode. 
We uncover an intrinsic relation between an odd micropolar modulus, the non-Hermitian skin effect, and a corresponding topological index. Numerical and experimental results show unidirectional amplification and attenuation of waves propagating through the metamaterial. Odd micropolarity extends the range of possible couplings between conventional strains/stresses and higher-order curvatures/moments by including antisymmetry in their relationship. The electronically assisted mechanical feedback provides an appealing solution to precisely modulate odd micropolar moduli without requiring changes to the metabeam's structure, geometry, or passive moduli. Our design can be flexibly tuned through computer coding and scaled via microelectromechancial systems (MEMS)~\cite{whitaker2018microelectronics,Cui2019Three}.
Our mechanical approach relies on a feedforward control loop, a generic concept that can exist in both metamaterial and biological contexts. The continuum theory also makes our approach especially generalizable to the mechanics in other systems.  Combining the principles illustrated here with disorder, nonlinearities, and strong dissipation suggests new approaches for the control of filaments and membranes arising in biological media~\cite{Needleman2017,Salbreux2017,prost2015}. 

\

\noindent{\bf Methods}
\setcounter{equation}{0}
\renewcommand{\theequation}{M\arabic{equation}}

\setcounter{figure}{0}
\renewcommand{\thefigure}{M\arabic{figure}}

\noindent{\textit{Sample Fabrication} \textemdash} 
The metabeam is composed of three piezoelectric patches (STEMiNC PZT 5J: $6~\text{mm}\times 4~\text{mm}\times 0.55~\text{mm}$) mounted via conductive epoxy onto a laser-cut stainless steal host beam. We achieve antisymmetric actuation without the use of an inverting voltage amplifier by mounting the two piezoelectric actuators such that their piezoelectric polarization directions are oppositely oriented.

\noindent{\textit{Experimental procedures} \textemdash} 
In experiments, nine metamaterial unit cells are connected with control circuits, see Figure~\ref{fig:Amp}a. Two piezoelectric transducers are attached on the left and right sides of the metamaterial to generate incident flexural waves. We employ ten-peak tone-burst signals with central frequencies ranging from 1.5 to 4.0 kHz in step sizes of 0.1 kHz. We generate and amplify incident wave signals via an arbitrary waveform generator (Tektronix AFG3022C) and a high voltage amplifier (Krohn-Hite), respectively. Transverse velocity wave fields are measured on the surface of the metamaterial by a scanning laser Doppler vibrometer (Polytec PSV-400). We note that the transfer-matrix method used to derive the theoretical curves in $\ref{fig:Amp}d,e$ rigorously assumes an infinite system. To experimentally approximate these conditions, we embed the active metamaterial within a larger host steal beam denoted by the grey region of Figure~\ref{fig:Amp}a. When waves cross the boundaries from host beam to the metamaterial, the reflection at boundaries between the host beam and the metamaterial is be negligible, as evidenced by our numerical and experimental results. To suppress reflected waves at the free boundaries of the host beam, we bonded two layers of clay on the host beam with sufficient lengths. This way, waves can propagate through the metamaterial with approximated infinite boundary conditions. The decay length is then obtained by calculating the wave amplitudes at different points in the metamaterial.

\noindent{\textit{Finite element simulations} \textemdash} 
We calibrate the transfer matrix method and continuum equations by conducting fully three-dimensional numerical simulations of the unit cell using the commercial finite element software COMSOL Multiphysics. In all the simulations, we model the piezoelectric patches via a three-dimensional linear piezoelectric constitutive law. The central piezoelectric patch acts as a sensor whose signal is obtained by integrating the free charge over the top surface the piezoelectric. 
The top and bottom surfaces of the piezoelectric sensor have zero electric potential. The bottom surfaces on the piezoelectric actuators are ground, and we apply electrical potentials on the top surfaces to act as actuating voltages. The actuating voltages are related to the sensing voltages via the electronic transfer function. 
For the wave dispersion computations in Fig.~\ref{fig:numer_main}a, Floquet periodic boundary conditions are applied on the left and right boundaries of a metamaterial unit cell. We calculate eigenfrequencies of the unit cell with different real wavenumbers. To simulate the wave propagation with open boundaries (Fi.g~\ref{fig:numer_main}b), a metabeam composed of 15 unit cells is placed between two external beams. Two perfectly matched layers (PMLs) are attached to both ends of the external beams in order to suppress reflected waves from the boundaries. The incident flexural wave is generated by applying a harmonic force on the boundary of the host beam. The out-of-plane displacement is measured at the left- and right-hand sides of the metabeam. The penetration depth is calculated by comparing the amplitudes of the two extracted displacements.

\noindent{\emph{Energy bounds on dynamic moduli} \textemdash} 
Here we discuss Eqs.~(\ref{eq:cond1}-\ref{eq:cond2}) in the main text. For simplicity, let us collect the stresses into a vector $\vb t = (\sigma_{zx}, M)^T$ and the deformations into a vector $\vb u = (s, b)^T $. Suppose the beam is subject to a deformation procedure such that its initial and final configurations at times $t = - \infty$ and $t= \infty$ are identical. Then the total work per unit volume done by the beam is given by 
\begin{align}
     \Delta W 
     =& \int_{-\infty}^\infty \dv{\vb u}{t} \cdot \vb t \dd t  \\
    =& - i \int_{- \infty}^\infty \omega  \vb u^\dagger (\omega) \cdot  \mathbf{C}(\omega) \cdot \vb u (\omega) \dd \omega   \\
    =&  \int_{0}^\infty \omega \vb u^\dagger (\omega) \cdot \mathbf{M}(\omega) \cdot \vb u(\omega) \dd \omega  \label{eq:w3}
\end{align}
In the final step, we have introduced the matrix $\mathbf{M}(\omega) = i [\mathbf{C}^\dagger(\omega) - \mathbf{C}(\omega)]$ and used the fact that $\vb u(-\omega)=\vb u^*(\omega)$ and $\mathbf{C}(-\omega)=\mathbf{C}^*(\omega)$. If the medium is passive, then we require that $\Delta W$ must be negative for all choices of $\vb u(\omega)$. Therefore, we require that the matrix $\mathbf{M}(\omega)$ be negative semidefinite. Using the parameterization in Eq.~(\ref{eq:param}), this requirement implies Eqs.~(\ref{eq:cond1}-\ref{eq:cond2}). See Refs.~~\cite{Srivastava2015causality,Day1971Time,Muhlestein2016Reciprocity} for related discussions in two and three dimensional media.

\noindent \emph{Cycles at finite frequency}\textemdash
 To gain intuition on the elastodynamics of the odd micropolar metabeam, it is useful to consider the notion of a cycle at finite frequency. At a finite frequency $\omega$, the modulus $P$ need not be real and we may write $P = \abs{P} e^{i \phi_P} $. 
In this case, both the real and imaginary parts of $P$ will contribute to the energy extracted. For example, consider a cyclic protocol which involves bending of amplitude $\abs{b}$ and shearing of amplitude $\abs{s}$ and relative phase delay $\phi_B$.  Applying Eq.~(\ref{eq:w3}), the total energy extracted per cycle is
\begin{align}
    \text{Work} = -\pi \abs{P}\abs{s} \abs{b} \sin(\phi_B + \phi_P) \label{eq:workfinite}
\end{align}
Notice that Eq.~(\ref{eq:workfinite}) gives mechanistic insight into the amplification of the waves observed in the experiment. 
When $P$ is nonzero, the eignemodes of $D(k)$ comprising a given plane wave will generically have a phase delay between bending and shearing. Hence, after one cycle, the nonconservative stresses will have converted stored electrical energy into mechanical energy. This conversion will cause the amplitude of the eigenvector to grow in proportion to its current amplitude, for which $\abs{s} \abs{b} $ is a proxy. Hence, the mode will be exponentially amplified, as reflected by the imaginary component of the eigenfrequencies.

\begin{figure}
    \centering
    \includegraphics[width=0.45\textwidth]{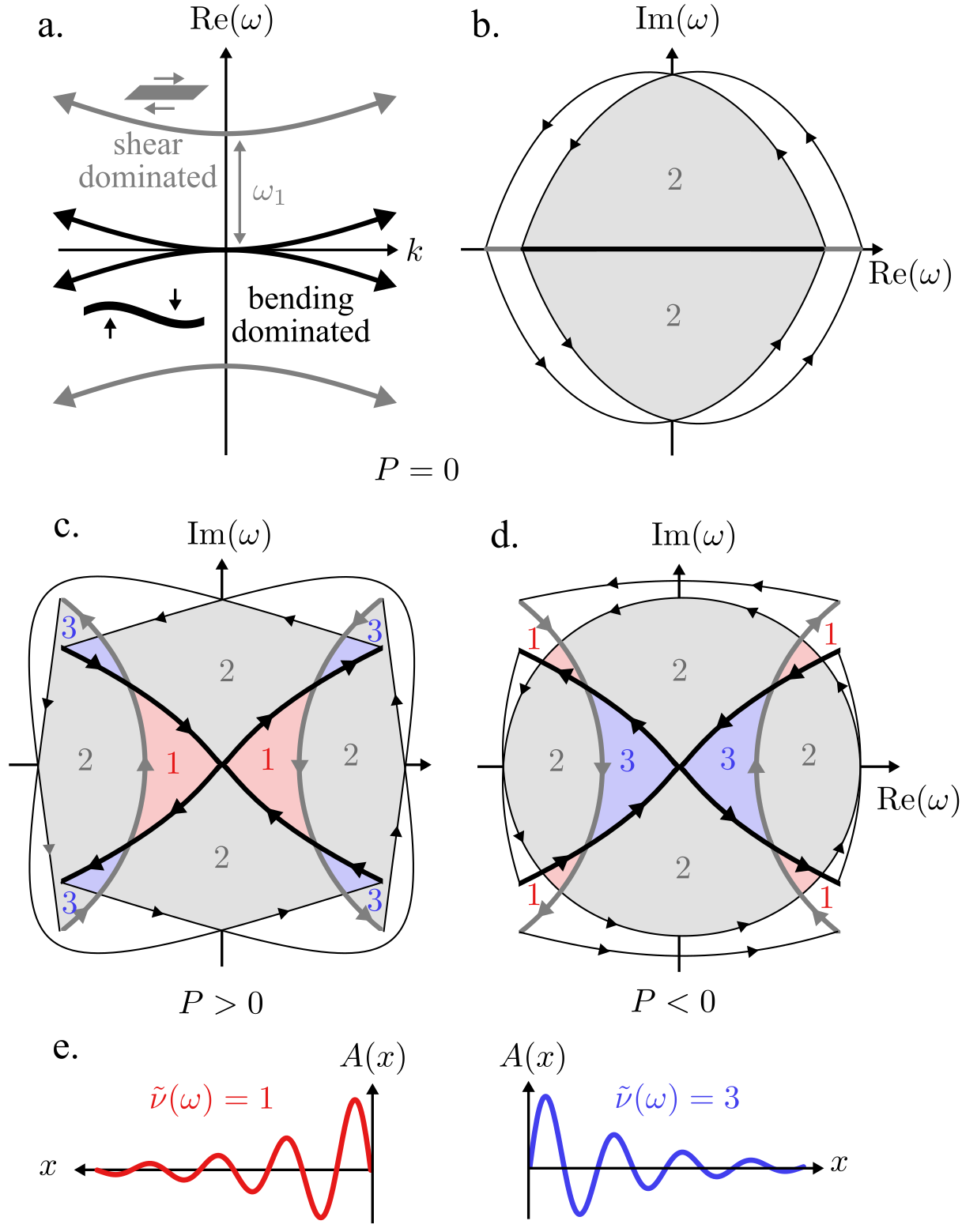}
    \caption{ {\bf Non-Hermitian band topology via odd micropolar elasticity.}~{\bf (a)}~The spectrum for $P=0$ features a pair of bending dominated bands (black) and shear dominated bands (grey) separated by a band gap $\omega_1$.~{\bf (b-d)}~The spectrum is shown in the complex plane $\omega$ plane for $P=0$, $P>0$, and $P<0$. The thick black lines represent the bending dominated band, while the thick grey lines represents the shear dominated bands, both with $k \in [-R,R]$ for a finite $R$. The thin black lines represent the analytical continuation of the spectrum for $k = R e^{i \phi}$ for $\phi \in [0 ,\pi]$. The arrows indicate the direction of increasing $k$. The numbers indicate the value of $\nu(\omega)$ for $\omega$ in the corresponding colored regions of the complex plane. This number corresponds to the number of times that the spectrum winds around a given region.~{\bf (e)}~For a semi-infinite system with a free boundary, the winding number of $\tilde \nu=1$ ($\tilde \nu=3$) for our continuum theory indicates a mode localized to the right (left) boundary. The wave forms schematically depict the localization with $A(x)$ representing amplitude. For a calculation of the precise eigenvectors, see S.I. Section S1F.} 
    \label{fig:contwind}
\end{figure}

\noindent \emph{Topological index in the continuum}\textemdash 
In this section, we discuss the index $\nu(\omega)$ from the point of view of the continuum theory~\cite{trefethen2020spectra}. 
The spectrum as a function of $k$ is plotted in Fig.~\ref{fig:contwind}a for $P=0$. The spectrum is given by the roots of the secular equation equation~(\ref{eq:sec}) and contains four roots since the equations of motion are second order in time and involve two coupled fields. 
We explicitly compute the eigenvectors and eigenvalues for small $k$ and $P=0$ in S.I. Section S1F.
The spectrum consists of a pair of Goldstone modes, which for small $k$ and $P=0$ represents a flexural motion of the beam. Additionally, the continuum equations imply an additional two modes separated by a band gap $\omega_1 \approx 10^5 \si{Hz} $. As shown in the S.I., for small $k$ and $P=0$, these modes are dominated by a shearing motion.
For the Goldstone mode, we expect a range of validity of the continuum theory at small $k$, since $\omega(k \to 0) = 0$. However, for the shear dominated mode, the continuum theory is not expected to self-consistently apply due to the finite gap. In S.I. Section S1G, we numerically compute the spectrum and eigenmodes for frequencies above the experimentally relevant range using COMSOL. There we also discuss why the winding number $\nu(\omega)$ computed in the continuum is still physically relevant despite the presence of additional high frequency modes not captured by the continuum. 

For a translationally invariant system, the difference between periodic and open boundary conditions is whether the differential operator that defines the equations of motion allows formal eigenvectors with complex wavenumber $k$. For a system with periodic boundary conditions, the spectrum consists of values of $\omega$ that solve $\det[ \dD(k) - \omega ]=0$ for real $k$. However, for a system on a semi-infinite domain $x \in [0, \infty)$, we allow $\Im(k) > 0$. These modes decay to the right as $\exp[(- \Im k + i \Re k )x ]$ and therefore maintain a finite $L^2$ norm, which represents the mechanical energy density. To determine whether a given frequency $\omega$ is in the semi-infinite boundary spectrum, we first must count the number of solutions to $\det[\dD(k) - \omega]=0$ with $\Im k >0$. We do so by applying Cauchy's argument principle from complex analysis:
\begin{align}
    \tilde  \nu(\omega) = \lim_{R \to \infty} \frac1{2\pi i}\oint_{\Gamma(R)} \dv{k} \log\det[ \dD(k) - \omega] \dd k, \label{eq:winding}
\end{align} 
where $\Gamma(R)$ is a counter clockwise curve in the complex plane given by $[-R,R]$ together with $R e^{i \phi}$ for $\phi \in [0, \pi]$. The winding number $\tilde \nu(\omega)$ itself does not directly determine the number of localized modes, since we must also consider the boundary conditions placed at $x=0$. Suppose that $\gamma$ independent homogeneous boundary conditions are placed at $x=0$.  Then the number of left localized modes will generically be given by:
\begin{align}
    \text{\# left localized modes at }\omega = \tilde \nu(\omega) - \gamma \label{eq:count1}
\end{align}
Likewise, for a system with boundary $( 0 , \infty]$, the number of right localized modes is given by 
\begin{align}
    \text{\# right localized modes at } \omega = d- \tilde \nu(\omega) -\gamma \label{eq:count2}
\end{align}
where $d$ is the degree of $\det[\dD(k) -\omega]$ as a polynomial in $k$. Whenever the right-hand side of Eq.~(\ref{eq:count1}) or Eq.~(\ref{eq:count2}) is negative, the mode count is taken to be zero.
In S.I. Section S1C, we provide a derivation of Eqs.~(\ref{eq:count1}) and Eq.~(\ref{eq:count2}) and explicitly define \emph{independent, homogeneous} boundary conditions. Such boundary conditions include, for example, stress-free ($M = \sigma_{zx} =0$) or motion-free ($\dot h = \dot \varphi =0$) boundaries.
Figure~\ref{fig:contwind}b-d illustrates the computation of the winding number. The thick black and grey lines are the periodic boundary spectrum for $k \in [- R, R]$, and the thin black lines are the analytical continuation for $k = R e^{i \phi}$ for $\phi \in [-\pi,\pi]$.  Colored regions are labeled by the value of $\tilde \nu(\omega)$. 
Suppose for example that the beam is given stress and moment free boundary conditions $\sigma_{zx}=0$ and $M =0$ at $x=0$. In this case $\gamma=2$, and therefore $\tilde \nu(\omega)=1$ indicates the presence of a right localized mode, $\tilde \nu(\omega)=3$ indicates the presence of a left localized modes, as shown in Fig.~\ref{fig:contwind}e. In the S.I. Section S1F, we explicitly compute examples of the eignenmodes in Fig.~\ref{fig:contwind}c.

\noindent \emph{Topological index from discrete models}\textemdash 
We now discuss the topological index in Eq.~(\ref{eq:wind1}), which is appropriate for discrete settings such as the discrete model in Section~\ref{sec:desc} and finite element simulations. 
Suppose the system is composed out of a unit cell of finite length $L$ whose internal state is represented by a vector $\Psib ( x)$. The components of $\Psib$ can represent, for example, the displacement and velocities of points in a finite element mesh. Here, $x$ is a discrete label that takes values in integer multiples of $L$. The Fourier transform of the equations of motion now read:
\begin{align}
\omega \Psib( k) = \DD(k ) \cdot \Psib (k) 
\end{align}
where $\DD(k)$ is the dynamical matrix for the discrete system, and $k$ is the wave number assuming values in $[-\pi/L, \pi/L]$. For a system with periodic boundaries, the spectrum is given by solutions to $\det[\DD(k) - \omega]=0$ for $k \in [- \pi /L, \pi/L]$.  For a system with semi-infinite boundaries (e.g. $x \in \{ 0, L ,2L, \dots \}$), we allow $\Im k >0$. As detailed in the S.I. Section S1E, we can invoke a similar application of the Cauchy argument principle to determine the number of eigenmodes at a given frequency $\omega$. To do so, we write $k=- i \log \lambda$, where $\lambda$ assumes values on the unit circle $S$ in the complex plane. We can then apply Cauchy's argument principle using $S$ as a counterclockwise contour for $\lambda$. We have:
\begin{align}
    \nu(\omega) =& \frac{1}{2\pi i}\int_S  \dv{\lambda} \log \det[\DD( -i \log \lambda) - \omega ] \dd \lambda \label{eq:topdes} \\
    =& \frac{1}{2\pi i}\int_{- \pi/L}^{\pi/L} \dv{k} \log \det[  \DD(k) - \omega ] \dd k  \\
    =& \frac{1}{2 \pi i } \sum_{\alpha } \int_{- \pi/L}^{\pi/L}  \dv{k} \log[\omega_\alpha (k) - \omega] \dd k 
\end{align}
where $\omega_\alpha(k)$ is the value of the periodic spectrum of the band $\alpha$, in agreement with  Eq.~(\ref{eq:wind1}) from the main text.

Notice that for the frequency denoted by the red star in Fig.~\ref{fig:numer_main}c, we have $\nu(\omega) = -1$. However, for the same point in Fig.~\ref{fig:contwind}c (which lies in the red region to the right of the origin), we have $\tilde \nu(\omega) =1$.
To see the relationship between these quantities, notice that $\tilde \nu (\omega)$ in Eq.~(\ref{eq:winding}) counts the number of zeros of $f(k)=\det[\dD(k) - \omega]$ in the upper half plane. 
However, in computing $\nu$ from Eq.~(\ref{eq:topdes}), the interior of $S$ contains not only the zeros of $F(\lambda)\equiv\det[ \dD(-i\log \lambda) - \omega]$ but also a set of poles. Each of these poles physically represents a boundary condition that arises when transitioning between a \emph{Laurent operator} to a \emph{Toeplitz operator} (see S.I. Section S1E for details). The number of poles depends on the precise discretization of the continuum equations. Hence the winding number $\nu(\omega)$ as given by Eq.~(\ref{eq:wind1}) represents the difference between the number of zeros (candidate modes) and the number of poles (boundary conditions). Therefore, one should compare $\nu(\omega)$ to $\tilde \nu (\omega) - \gamma$ for an appropriate choice of $\gamma$. In practice, $\gamma$ is determined by physically interpreting the boundary conditions implied by the discretization. In S.I. Section S1E, we show that the discretization used in the discrete model of Section~\ref{sec:desc} and the COMSOL simulations enforce displacement-free boundary conditions and hence corresponds to $\gamma =2$. Hence, $\tilde \nu(\omega) - \gamma = \nu (\omega) =-1 $ as required for physical consistency.

\noindent \emph{Calculation of penetration depth\textemdash} In Figure~\ref{fig:numer_main}b, we show the penetration depth for modes associated with positive real $\omega$. We can compute this expression by solving $\omega_+(k)=\omega$ in Eq.~(\ref{eq:approx1}) for complex $k$. In particular, substituting  $k = q + i \kappa$ into Eq.~(\ref{eq:approx1}) yields:
\begin{align}
    0=&2 k \kappa + \frac{P}{2 \mu} (k^3 - 3 k \kappa^2) \label{eq:k1} \\
    \omega=&\sqrt\frac{B}{\rho} \qty[\qty(1- \frac{P \kappa}{\mu}) \qty(k^2- \kappa^2)  - \frac{2P}{\mu} k^2 \kappa ] \label{eq:k2}
\end{align}
Solving Eqs.~(\ref{eq:k1}-\ref{eq:k2}) to leading order in $k$, we obtain Eq.~(\ref{eq:kappa}).

\

\noindent{\bf Acknowledgements}
The authors gratefully thank Prof. Hussein Nassar from University of Missouri for valuable discussions, and Michele Fossati for a critical reading of the manuscript. 
This work is supported by the Air Force Office of Scientific Research under Grant No. AF 9550-18-1-0342 and AF 9550-20-0279 with Program Manager Dr. Byung-Lip (Les) Lee and the Army Research Office under Grant No. W911NF-18-1-0031 with Program Manager Dr. Daniel P Cole. V.V.~was supported by the Complex Dynamics and
Systems Program of the Army Research Office under grant W911NF-19-1-0268. C.S.~was supported by the National Science Foundation Graduate Research Fellowship under Grant No.~1746045. Some of us benefited from participation in the KITP program on Symmetry, Thermodynamics and Topology in Active Matter supported by Grant No. NSF PHY-1748958.

\

\noindent{\bf Author contributions}
Y.C. and G.H. conceived the concept; X.L. conducted experiments; Y.C. and C.S. performed theoretical investigations; Y.C. performed numerical investigations; G.H. and V.V. supervised the research; All the authors discussed the results; all authors wrote the manuscript and interpreted the results. 

\

\noindent{\bf Competing interests}
The authors declare no competing interests.

\

\noindent{\bf Data and materials availability}
All data needed to evaluate the conclusions in the paper are present in the paper and/or the Supplementary Information. Extended data, software, and materials in the main text and the Supplementary Information are available upon request by contacting the corresponding authors.

\

\clearpage

\onecolumngrid

\begin{center}
{\bf \large Supplementary Information}
\end{center}

\renewcommand{\theequation}{S\arabic{equation}}
\setcounter{equation}{0}
\renewcommand{\thetable}{S\arabic{table}}  
\renewcommand{\thefigure}{S\arabic{figure}}
\setcounter{figure}{0}
\renewcommand\figurename{Fig.}
\renewcommand{\thesection}{S\arabic{section}}
\setcounter{section}{0}

\section{Continuum theory}

\subsection{Numerical characterization of elastic moduli}

Our characterization of the elatic moduli begins with the standard kinematic assumptions of Timoshenko beam theory~\cite{Timoshenko1949}.
Let $u_x(x,y,z)$, $u_y(x,y,z)$, and $u_z(x,y,z)$ be the underlying continuum displacement field. Here, the $x$ direction is oriented along the beam, the $y$ direction is oriented out of plane, and the $z$ direction is oriented vertically. In following Timoshenko beam theory, we assume: 
\begin{enumerate}
    \item There are no out-of-plane displacements: $u_y(x,y,z) =0$.
    \item Cross sections of the beam remain planar. Hence, we may introduce the following parameterization:
    \begin{align}
        u_z(x,y,0) =& h(x) \\
        u_x(x,y,z) =& -z \varphi(x) + u(x), \label{eq:var}
    \end{align}
    where $z=0$ coincides with the midplane of the beam. 
\end{enumerate}
Under these assumptions, $h(x)$, $\varphi(x)$, and $u(x)$ are the independent degrees of freedom in our effective 1D model. In the main text, we assume that $u(x) =0$, as is valid in experiments. Nonetheless, we can probe elongations of the beam in simulation, and hence we include $u(x)$ in our analysis here. The equations of motion are obtained by invoking conservation of angular momentum in the $\hat y$ direction, linear momentum in the $\hat x$ direction, and linear momentum in the $\hat z$ direction. These conserved quantities give rise to the following equations of motion:
\begin{align}
       I \ddot \varphi =& \partial_x M + \sigma_{zx} \\
        \rho \ddot h =& \partial_x \sigma_{zx} \\
    \rho \ddot u =& \partial_x \sigma_{xx},
\end{align}
where $\rho$ is the volumetric mass density, $I$ is the cross sectional moment of inertia, $M$ is the bending moment, and $\sigma_{ij}$ is the stress tensor. 
Next, we expand the moments and stresses to linear order in the the deformations of the beam, which include bend $\partial_x \varphi$, shear $\partial_x h - \varphi$, and elongation $\partial_x u$. We can summarize the linear response by a 3 by 3 matrix $C_{ij}$:
\begin{align}
    \mqty(  \sigma_{zx}  \\ M \\ \sigma_{xx}  ) = \mqty ( C_{11} & C_{12} & C_{13} \\
    C_{21} & C_{22} & C_{23} \\
    C_{31} & C_{32} & C_{33}
    ) \label{eq:const}
    \mqty( \partial_x h - \varphi \\ \partial_x \varphi  \\   \partial_x u ).
\end{align}
Eq.~(\ref{eq:const}) is an enlarged version of Eq.~(10) in the main text. 
It is useful first to  make predictions on which moduli will dominate Eq.~(\ref{eq:const}) based on the properties of the microscopic unit cell presented in Fig.~1 of the main text. First, we expect the beam to inherit the Young's modulus, $E$, shear modulus $\mu$, and bending moment $B$ present in standard beam theory. Moreover, since elongation or compression of the central piezoelectric sensor induces antisymmetric stresses in the piezoelectric actuator, we expect elongation $\partial_x u$ and bending $\partial_x \varphi $ to give rise to shear stress $\sigma_{zx}$. Hence, we expect $C_{ij}$ to take the form:
\begin{align}
C_{ij}= \mqty(\mu & P & K \\ 0 & B & 0 \\ 0 & 0 & E) \label{cmat}
\end{align}
In Eq.~(\ref{cmat}), the moduli $K$ and $P$ are introduced via the piezoelectric feedback. The modulus $P$ is the odd micropolar modulus, which is the primary focus of this work. The modulus $K$ is an asymmetric modulus between shear and elongation. 

\begin{figure}
    \centering
    \includegraphics[width=0.75\textwidth]{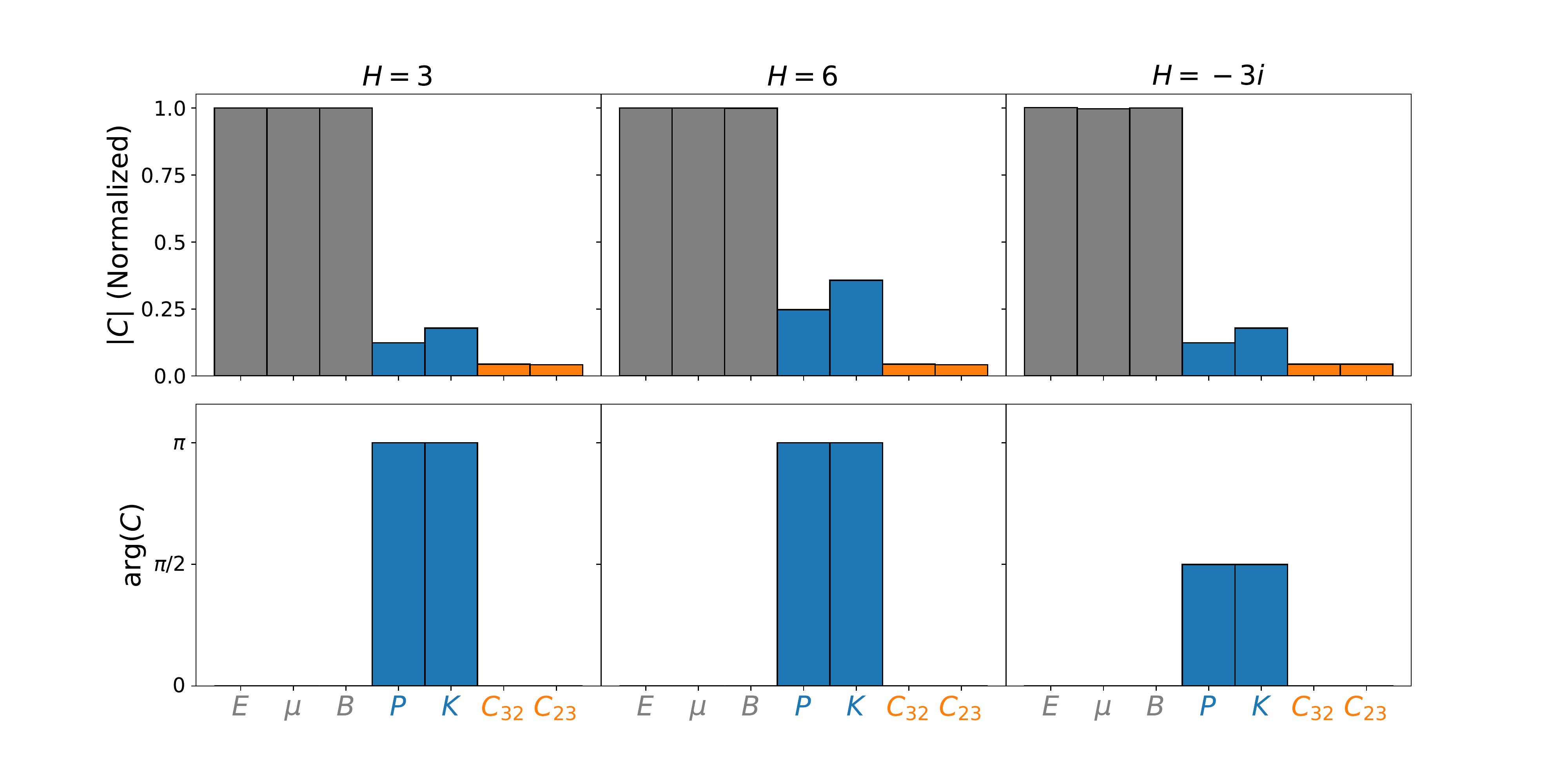}
    \caption{Determination of linear response via finite-element simulation. The normalized magnitude and complex argument of the nonzero moduli are shown for three different values of the transfer function $H$. The quantities $E$, $\mu$, and $B$ are normalized by their passive values $E_0$, $\mu_0$, $B_0$ (determined when $H=0$). The modulus $K$ is normalized by $\mu_0$. The moduli $P$, $C_{23}$, and $C_{32}$ are normalized by $\mu_0 h_b$, where $h_b$ is the thickness of the metabeam (see section \ref{expt}). }
    \label{fig:Modulus}
\end{figure}

Supplementary Figure~\ref{fig:Modulus} shows the result of finite-element simulations  that directly probe the response of the metabeam. The moduli are determined by applying strain controlled boundary conditions at the terminating faces of a single unit cell and computing the reaction forces. We find that $E=152.4 \times 10^9 \si{ kg/m s^2}$, $\mu=1.3 \times 10^9 \si{kg / m s^2}$, and $B=0.112 \times 10^6 \si{kg/s^2}$ are independent of the transfer function $H(\omega)$. Moreover, we confirm that when the electronic feedback is present, the linear response $\mathbf{C}$ contains nonzero $K$ and $P$. We empirically find that $K$ and $P$ are proportional to the transfer function $H(\omega)$ via the relationship $K= \Lambda H $ and $P = \Pi H$, for (real) material constants $\Lambda=1.2 \times 10^9 \si{kg/ m s^2}$ and $\Pi=4.7 \times 10^6 \si{kg m/ s^2}$. Additionally, we find small, but non-zero, values for $C_{23}$ and $C_{32}$. This coupling between elongation and bending arises since the piezoelectrics are mounted only on the top surface of the metabeam. However, since  $\abs{C_{32}}$ and $\abs{C_{23}}$ are small, we can safely neglect these terms in the subsequent analysis. 

\begin{figure}
    \centering
    \includegraphics[width=0.7\textwidth]{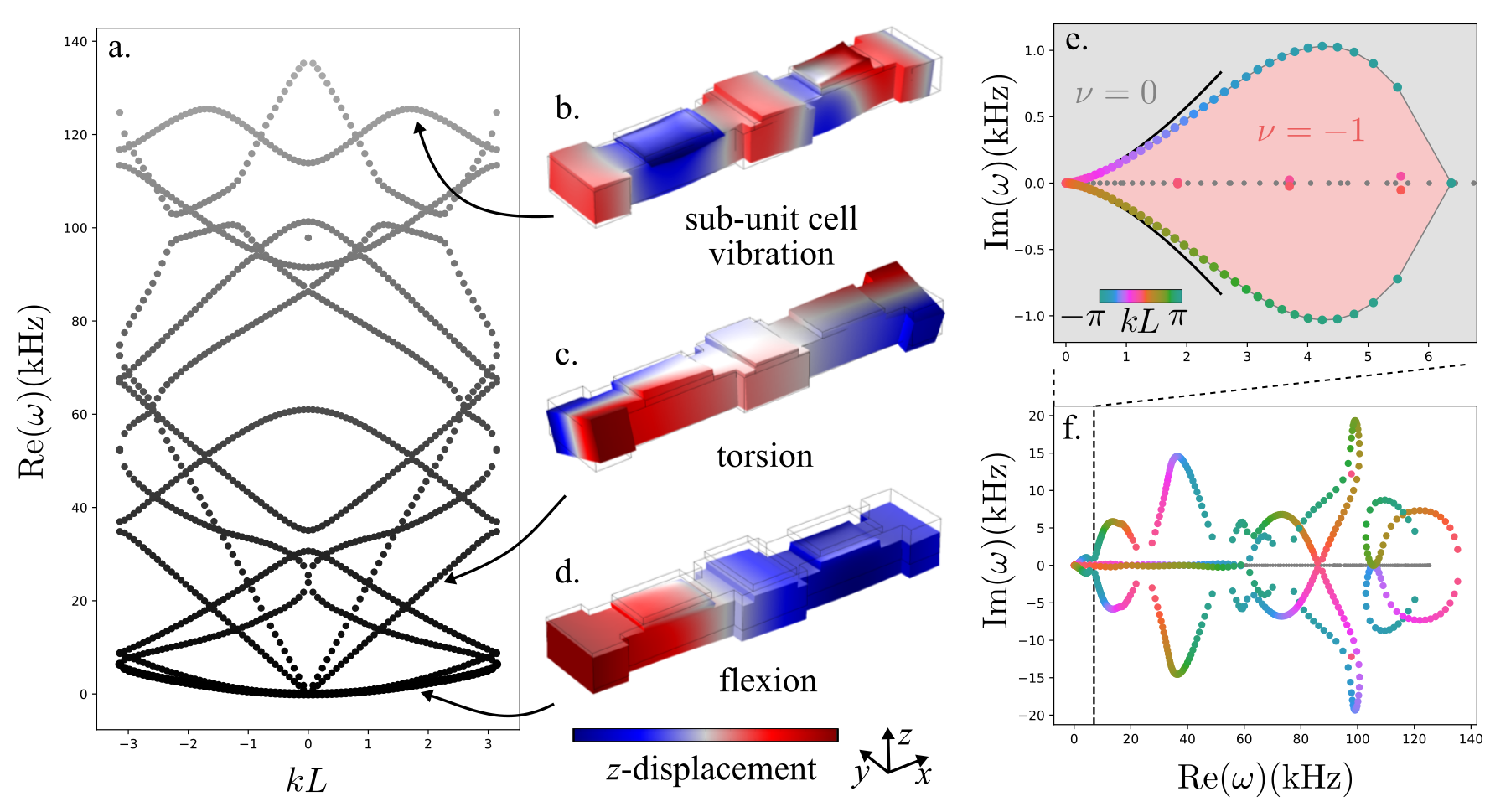}
    \caption{{\bf (a)} The fifteen lowest eigenfrequencies computed in COMSOL for the active metabeam with $H(\omega) = 3$, and hence $P = 3 \Pi$.  {\bf (b-d)}  The eigenmodes associated with selected branches. The color indicates the vertical displacement.
    { \bf (e)} The vibrational spectrum with periodic boundary conditions for low frequencies (${\lesssim}7 \si{kHz}$) are plotted in the complex $\omega$ plane. The color of the data points indicates the wave number $kL$, where $L$ is the length of a unit cell. The solid black lines indicate the continuum theory. For simplicity, only the vibrational frequencies with positive real part are shown. The red region has a winding number of $\nu=-1$ and the grey region has a winding number of $\nu=0$, as determined by Eq.~(30).  Numerical modes with negligible imaginary part are represented by the small grey dots. For these simulations, we set the transfer function to $H(\omega)=3$ and hence $P = 3 \Pi$. 
    { \bf (f) } The spectrum for the lowest 15 modes in the complex $\omega$, illustrating the complexity of the full spectrum of the metabeam. Same color scheme as in panel (a). 
    }
    \label{fig:high_freq}
\end{figure}

\begin{figure}
    \centering
    \includegraphics[width=0.75 \textwidth]{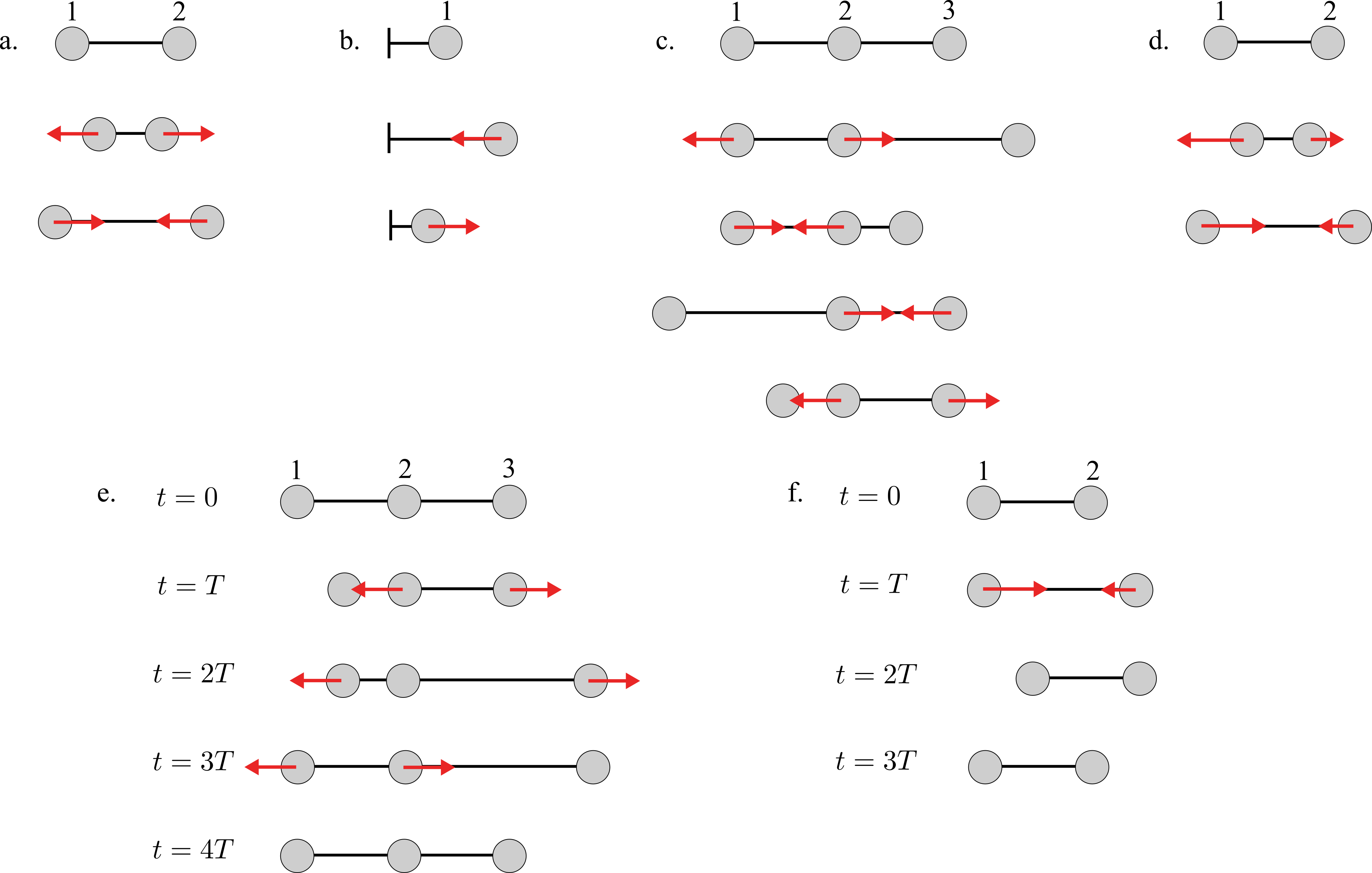}
    \caption{  {\bf Illustrations of reciprocity violation.}~Four model systems are shown in which masses are connected by Hookean springs, or generalizations of Hookean springs. The top row shows the rest state of each system, while deformed states are shown beneath. The red arrows indicate forces. 
    {\bf a.}~A Hookean spring exerts equal and opposite forces on the two attached particles when stretched or compressed. This system respects linear momentum conservation and Maxwell-Betti reciprocity. {\bf b.}~A mass is connected to a fixed wall by a Hookean spring. Due to the presence of the wall, the linear momentum of the mobile particle is not conserved. Nonetheless, Maxwell-Betti reciprocity is respected. {\bf c.}~Three masses are connected by generalized springs.   When the right bond (connecting particles 2 and 3) is stretched/compressed, an outward/inward tension is exerted on the left bond (connecting particles 1 and 2). However, when the left bond is stretched/compressed, an inward/outward tension is exerted on the right bond. In each deformed state, the forces on all particles sum to zero, indicating that linear momentum is conserved. Nonetheless, the system violates Maxwell-Betti reciprocity due to the asymmetry in the relationship between the right and left bond. Such a system can be realized via sensors and linear actuators coupled by electronic feedback. Such a system requires an energy source, but no external medium to provide the linear momentum. 
    {\bf d.}~A generalized spring is constructed in which the forces acting on the two masses are not equal and opposite. Such a device requires an external source of linear momentum (not shown) to provide the force imbalance, and a source of energy to compensate for the violation of Maxwell-Betti reciprocity. {\bf e-f.} Energy extracting cycles performed with the systems in panels c. and d., respectively.   }
    \label{fig:toy}
\end{figure}

\subsection{Notions of reciprocity}
Here we provide a list of four minimal systems that illustrate the distinction between reciprocity in the sense of momentum conservation and Maxwell-Betti reciprocity. For simplicity, we consider one-dimensional systems for which linear momentum is the relevant form of momentum. 
 
\

\noindent {\bf System 1}  The first system is a simple Hookean spring connecting two masses, see Fig.~\ref{fig:toy}a. This system respects both linear momentum conservation and  Maxwell-Betti reciprocity. The system has two degrees of freedom $u_1$ and $u_2$ being the displacements of the two particles. There are two conjugate forces $F_1$ and $F_2$ that are defined via stipulation of the virtual work:
    \begin{align}
        \dd W = F_1 \dd u_1 + F_2 \dd u_2
    \end{align}
    The system comes with a linear constitutive relation:
    \begin{align}
        \mqty[  F_1 \\ F_2 ] =  \underbrace{\mqty[ -k & k \\ k & -k  ]}_{\Cb} \mqty[u_1 \\ u_2 ]  \label{eq:hookf} 
    \end{align}
    where $k$ is the spring constant. Since the linear response matrix $\Cb$ is symmetric, the linear response is compatible with a quadratic potential energy $V = \frac{k}2 (u_1- u_2)^2$ and hence respects Maxwell-Betti reciprocity. Secondly, we assume dynamic forms $ \dv{p_1}{t} = F_1$ and $\dv{p_2}{t} =F_2$ where $p_1$ and $p_2$ are the momenta associated with masses $1$ and $2$. The rate of change of the total momentum $P = p_1 + p_2$ is given by:
    \begin{align}
        \dv{P}{t}= \dv{p_1}{t}  + \dv{p_2}{t} = F_1 + F_2 = 0 
    \end{align}
    where we have used Eq.~(\ref{eq:hookf}).

\

\noindent {\bf System 2 }  The second system we present respects Maxwell-Betti reciprocity, but does not conserve linear momentum. This system is a mass pinned to a substrate by a Hookean spring, see Fig.~\ref{fig:toy}b. This system is described by a single degree of freedom $u_1$. There is one conjugate force $\dd W = F_1 \dd u_1$ that is given by 
    \begin{align}
    F_1 = k u_1
    \end{align}
In this case, the linear response matrix $\Cb$ is simply a scalar $\Cb=k$. Thus we have $\Cb=\Cb^T$ and the system obeys Maxwell-Betti reciprocity and it is compatible with an energy function $V = \frac 12 u_1^2$. However, the total linear momentum in the system $P=p_1$ is manifestly not conserved since there are configurations of the system for which $\dv{P}{t} = F_1$ is nonzero. Indeed, linear momentum is conserved if the substrate is included in the analysis. However, we take the point of view that the substrate is not dynamical (since perhaps it is very large) and hence a more reasonable choice of system is the single mass.

\

\noindent  {\bf System 3}  Now we provide an example of a system in which linear momentum is conserved but Maxwell-Betti reciprocity is violated. As shown in Fig.~\ref{fig:toy}c, consider a collection of three coupled masses with three degrees of freedom $u_1$, $u_2$, and $u_3$. The the system has three conjugate forces defined through a statement of virtual work $\dd W = F_1 \dd u_1+ F_2 \dd u_2 + F_3 \dd u_3$. The masses are coupled through specially designed actuators such that they have the following constitutive relationship
    \begin{align}
        \mqty[ F_1 \\ F_2 \\ F_3 ] = \underbrace{\mqty[ 0 & k^a & -k^a \\ -k^a & 0 & k^a \\ k^a &  -k^a & 0    ]}_{\Cb} \mqty[ u_1 \\ u_2 \\ u_3] 
    \end{align}
    In this case, the linear response matrix $\Cb$ is not symmetric. This is an indication that the system violates Maxwell-Betti reciprocity. To see that an energy function cannot be defined, consider the following protocol illustrated in Fig.~\ref{fig:toy}e.
    \begin{align}
        u_1(t) =& 
        \begin{cases}
        t U & t \in [0,T]  \\
         T U & t \in [T , 2T ] \\
        U (3T -t)  & t \in [2T, 3T] \\
        0 & t \in [3T, 4T]. 
        \end{cases} \\
        u_2(t) =& 0 \\
        u_3(t) =&  \begin{cases}
        0 & t \in [0,T]  \\
         (t-T) U & t \in [T , 2T ] \\
        U T  & t \in [2T, 3T] \\
        U (4T-t)  & t \in [3T, 4T].
    \end{cases} 
    \end{align}
    The work done along this closed cycle in the configuration space can be easily computed to be $\int_0^{4T} \dd W = \frac12 U^2 k^a$. Since the work is nonzero along a closed cycle, a potential energy function cannot be defined. Nonetheless, the total linear momentum  $P=p_1 + p_2 + p_3$ is conserved since:
    \begin{align}
        \dv{P}{t} = F_1 + F_2 + F_3 =0
    \end{align}
    for all configurations.

\

\noindent {\bf System 4 } Finally, we consider an example that violates both Maxwell-Betti reciprocity and linear momentum conservation~\cite{Brandenbourger2019}. The model has two degrees of freedom $u_1$ and $u_2$ and two conjugate forces $F_1$ and $F_2$.
The constitutive relation is given by:
    \begin{align}
        \mqty[F_1 \\ F_2 ] = \underbrace{ \mqty[ -k(1+\epsilon) & k (1+\epsilon) \\ k(1-\epsilon) & -k (1- \epsilon ) ]}_{\Cb} \mqty[ u_1 \\ u_2]
    \end{align}
    Here, the linear response matrix is no longer symmetric $\Cb \neq \Cb^T$, indicating that Maxwell-Betti reciprocity is violated. Moreover, as illustrated in Fig.~\ref{fig:toy}f, one can perform a cycle in the configuration space that extracts work: 
        \begin{align}
        u_1(t) =& 
        \begin{cases}
        0 & t \in [0,T]  \\
         (t-T) U & t \in [T , 2T ] \\
          UT  & t \in [2T,3T] 
        \end{cases} \\
        u_2(t) =&  \begin{cases}
        U t & t \in [0,T]  \\
         T U & t \in [T , 2T ] \\
        U (3T-t)  & t \in [2T, 3T] 
    \end{cases} 
    \end{align}
    The total work done is $\int_0^{3T} \dd W = 2 k \epsilon  U^2 $. However,  the total linear momentum $P = p_1+ p_2$ is not conserved since
    \begin{align}
        \dv{P}{t} = -  k \epsilon (u_1 -u_2)
    \end{align}
    which is nonvanishing. To physically realize such a system, a linear momentum sink must be present in addition to a source of energy to provide the Maxwell-Betti reciprocity violation.

\subsection{Dispersion relations calculated using the discrete model}
Fig. S4 shows a comparison of the dispersion relations for the discrete model and  continuum theory.\

\begin{figure}
    \centering
    \includegraphics[width=0.4\textwidth]{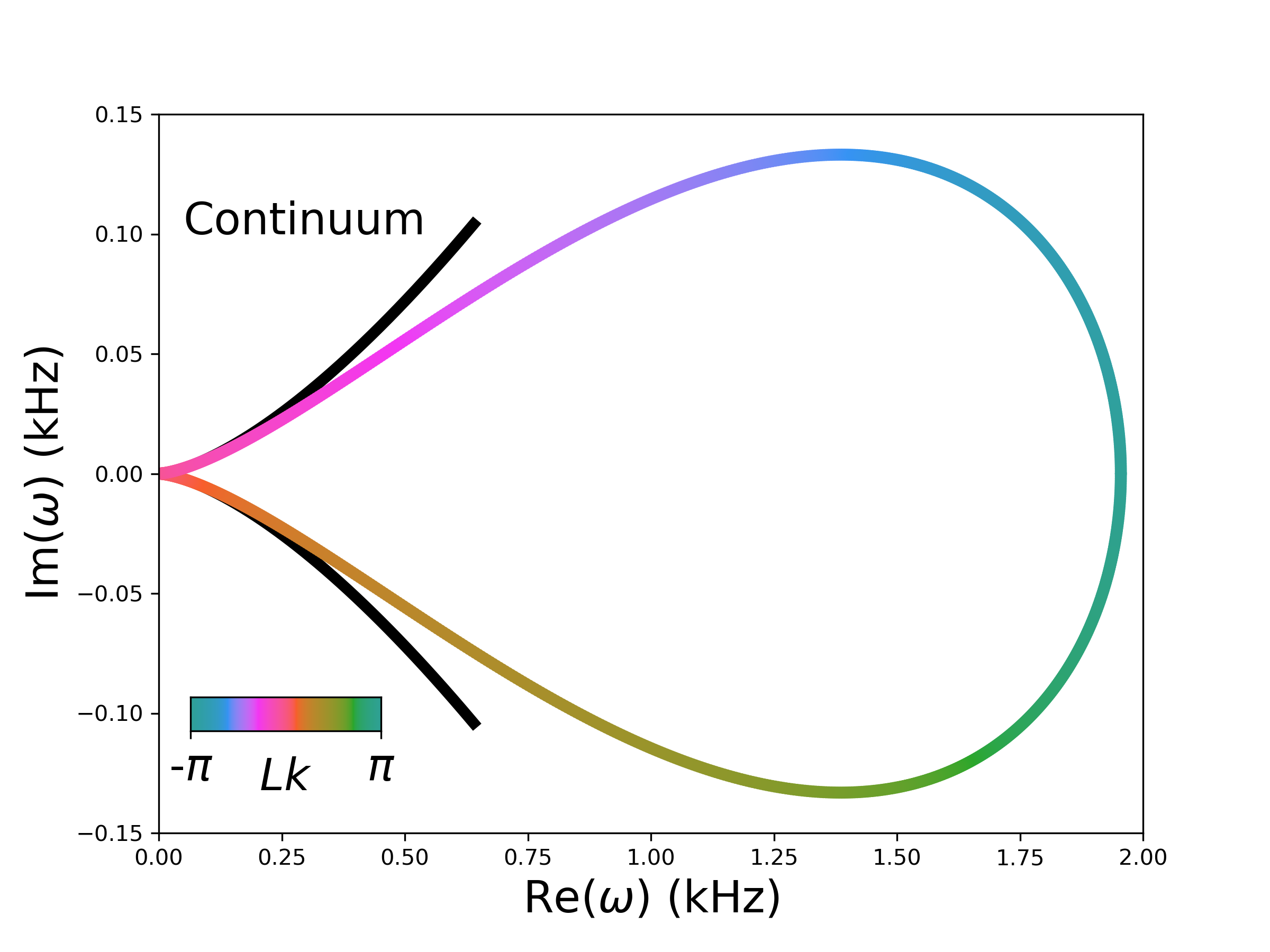}
    \caption{  {\bf Dispersion relations of the discrete model of the odd micropolar metabeam calculated}. Here we use for $p>0$ (parameters are the same as those in Fig. 3a), and the color bar represents the normalized wavenumber $kL$. See Section IIF of the main text for the description of the model. The solid black line represents the continuum theory with wave number $kL \in [-0.96,0.96]$.  }
    \label{fig:Disc_model_dsps}
\end{figure}

\begin{figure}
    \centering
    \includegraphics{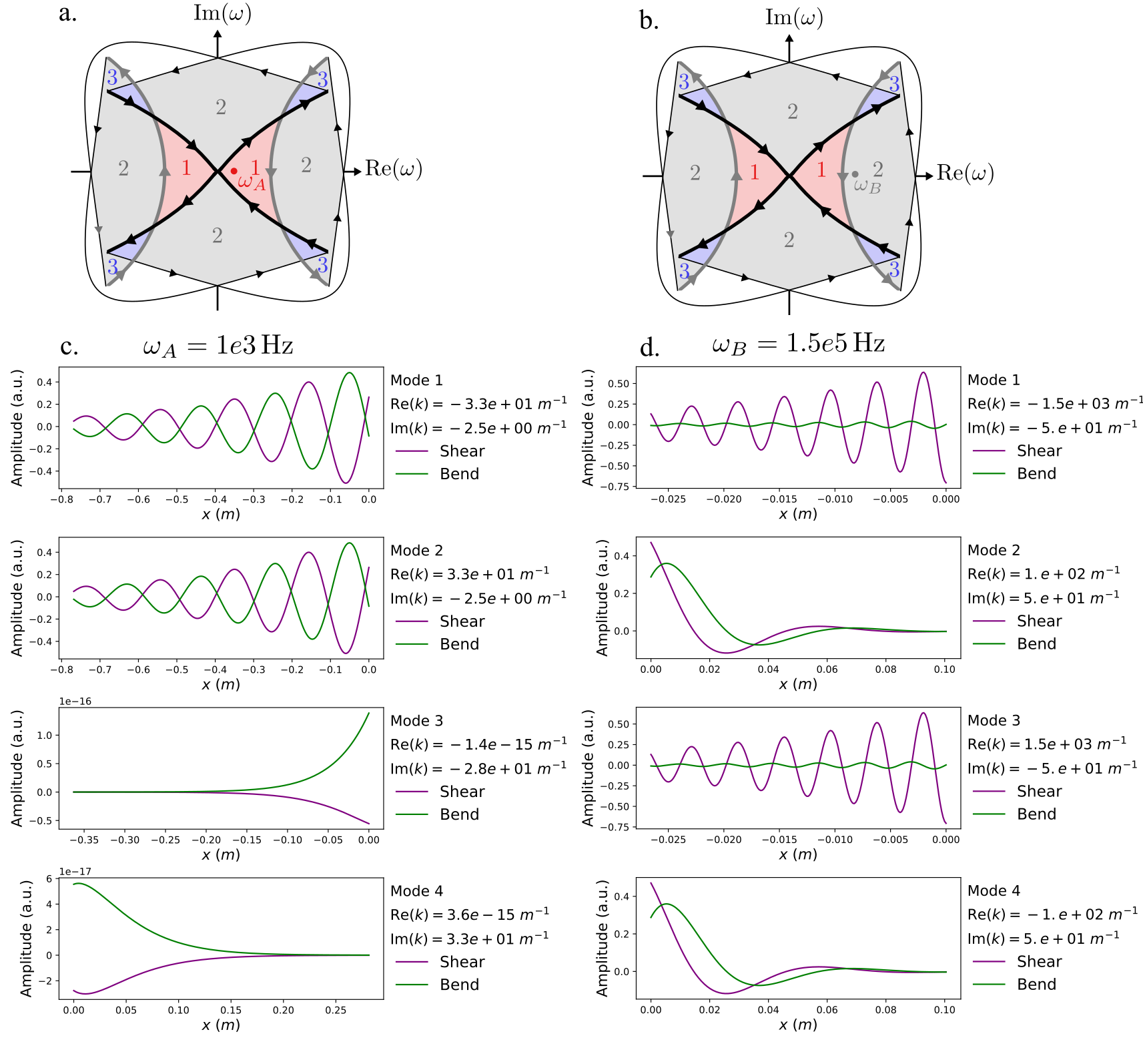}
    \caption{  {\bf Illustration of localized modes from continuum theory.} {\bf a-b.} The spectrum in the complex plane for $P>0$, using notation identical to that used in Figure~M1. Here we highlight two modes $\omega_A = 1\times10^{3} \si{Hz}$ and $\omega_{B} = 1.5 \times 10^5 \si{Hz} $, schematically placed within the complex plane.  We compute the eigenmodes of Eq.~(\ref{eq:dmat2}) at frequencies $\omega_A$ and $\omega_B$ using the parameter $\omega_1 = 10^5 \si{Hz}$, $\ell_1 = 10^{-3} m$, $\ell_2 = 10^{-2} m$, $\tilde P=1$. In {\bf c-d.}~we show the shearing $\tilde s$ and bending $\tilde b$ profiles over space. Consistent with the winding number $\tilde \nu(\omega_A) =1$, there is one left localized mode and three right localized modes at $\omega_A$. Moreover, we have $\tilde \nu(\omega_B) =2$, which is consistent with two left-localized modes and two right-localized modes. 
    }
    \label{fig:modes}
\end{figure}

\subsection{Boundary conditions in the continuum}
Here we derive the index Eqs.~(M6) and (M7) in the main text. For concreteness, we will perform the derivation in the context of the continuum theory for the active metabeam, with sufficient discussion to allow generalization. For further discussion, see Ref.~\cite{trefethen2020spectra} and references therein. 
Consider a system with semi-infinite boundary conditions $[0,\infty)$ governed by the dynamical matrix $\dD(k)$. In the notation of Eq.~(23), we seek to determine many vectors $\Psib =(\tilde p_h, \tilde p_\varphi \tilde s , \tilde b) $ satisfy the equation
\begin{align}
[\dD(k) - \omega ] \cdot \Psib =0    
\end{align}
with $\Im(k) > 0$ subject to $\gamma$ \emph{independent}, \emph{homogeneous} boundary conditions at $x=0$. By independent, homogeneous boundary conditions, we mean boundary conditions that can be stated in the following form:
\begin{align}
\begin{bmatrix}
B_{1,1} &    \dots & B_{1,4(N+1) } \\ 
\vdots & &   \vdots \\
B_{\gamma, 1}  & \dots &  B_{\gamma,4 (N+1)}  
\end{bmatrix}
\begin{bmatrix}
\tilde p_h(x) \\
\tilde p_\varphi (x) \\
\tilde s (x) \\
\tilde b (x) \\
\vdots \\
\partial_x^{N} \tilde p_h (x)  \\
\partial_x^{N} \tilde p_\varphi (x) \\
\partial_x^{N} \tilde s (x) \\
\partial_x^{N} \tilde b (x)
\end{bmatrix}_{x=0} =0 
\end{align}
where $\mathbf{B}$ is a matrix of rank $\gamma$. To count the number of solutions to $[ \dD(k) - \omega ] \cdot \Psib=0$, we introduce the winding number
\begin{align}
    \tilde \nu(\omega) = \lim_{R \to \infty} \frac1{2\pi i} \oint_{\Gamma(R)} \dv{k} \log f(k) \dd k
\end{align}
where $f(k) = \det [ \dD(k) - \omega]$ and $\Gamma(R)$ is the contour $[-R,R]$ completed by $Re^{i\phi}$ with $\phi\in[0,\pi]$. Notice that $f(k)$ is a polynomial with non-negative powers of $k$. Let $d$ be the rank of the polynomial $f(k)$. To simplify the derivation below, we will assume without loss of generality that $4 N \ge d$. This is not a restriction because it can always be achieved by augmenting $\mathbf{B}$ with zero columns. Since $f(k)$ contains only positive powers of $k$, the poles of $f(k)$ are located at $\abs{k} \to \infty$ and hence are not included in the contour $\Gamma(R)$ for any $R$. Thus, by Cauchy's argument principle, $\tilde \nu(\omega)$ counts the number of zeros of $f(k)$ with $\Im(k) > 0$. Each zero $k_a$ has an associated ``candidate" eigenvector $\Psi^a$, with $a =1, 2, \dots, \tilde \nu$. Therefore, we may write:
\begin{align}
\begin{bmatrix}
\tilde p_h (x) \\
\tilde p_\varphi (x) \\
\tilde s (x) \\
\tilde b (x)  \\
\vdots \\
\partial_x^N \tilde p_h (x)  \\
\partial_x^N \tilde p_\varphi (x) \\
\partial_x^N \tilde s (x) \\
\partial_x^N \tilde b (x) 
\end{bmatrix}_{x=0} 
=
\underbrace{ 
\begin{bmatrix}
\tilde p_h^1 & \cdots & \tilde p_h^{\tilde \nu}  \\
\tilde p_\varphi^1 & \cdots & \tilde p_\varphi^{\tilde \nu}  \\
\tilde s^1 & \cdots & \tilde s^{\tilde \nu} \\
\tilde b^1 & \cdots & \tilde b^{\tilde \nu} \\
\vdots \\
k^N_1 \tilde p_h^1 & \cdots &  k^N_{\tilde \nu} \tilde p_h^{\tilde \nu} \\
k^N_1 \tilde p_\varphi^1 & \cdots & k^N_{\tilde \nu} \tilde p_\varphi^{\tilde \nu}  \\
k^N_1 \tilde s^1 & \cdots & k^N_{\tilde \nu} \tilde s^{\tilde \nu}  \\
k^N_1 \tilde b^1 & \cdots & k^N_{\tilde \nu} \tilde b^{\tilde \nu}
\end{bmatrix}}_{ \mathbf{H} }
\begin{bmatrix}
g_1 \\
g_2 \\
\vdots \\
g_{\tilde \nu} 
\end{bmatrix}
\end{align}
We seek to determine the number of linearly independent vectors $\mathbf{g}$ that satisfy the boundary conditions $ \mathbf{B} \cdot \mathbf{H} \cdot  \mathbf{g} = 0$. Thus, we seek to determine $\dim \ker \mathbf{B} \cdot  \mathbf{H} $. 
Since $4N \ge d \ge  \tilde \nu $, the columns of $\mathbf H$ are linearly independent and hence $\rank \mathbf{H}  =\tilde \nu $. 
Therefore, we may write:
\begin{align}
    \mathbf{B} =& \mqty[  \mathbf{a}_1 & \mathbf{a}_2 & \cdots & \mathbf{a}_\gamma ] \mqty[  \mathbf{b}_1^T \\  \mathbf{b}_2^T \\ \vdots \\  \mathbf{b}_\gamma^T ] \\
        \mathbf{H} =& \mqty[  \mathbf{c}_1 & \mathbf{c}_2 & \cdots & \mathbf{c}_{\tilde \nu}  ] \mqty[  \mathbf{d}_1^T \\  \mathbf{d}_2^T \\ \vdots \\  \mathbf{d}_{\tilde \nu} ^T ] \\
\end{align}
where each set of vectors $\{ \mathbf{a}_i \} $, $\{ \mathbf{b}_i \} $, $\{ \mathbf{c}_i \} $, $\{ \mathbf{d}_i \} $ are individually linearly independent. Therefore, we may write
\begin{align}
    \mathbf{B} \cdot \mathbf{H} = \mqty[  \mathbf{a}_1 & \mathbf{a}_2 & \cdots & \mathbf{a}_\gamma ]  
    \underbrace{ \mqty[ \mathbf{b}_1 \cdot \mathbf{c}_1 & \mathbf{b}_1 \cdot \mathbf{c}_2 & \cdots &  \mathbf{b}_1 \cdot \mathbf{c}_{\tilde \nu}\\
                 \mathbf{b}_2 \cdot \mathbf{c}_1 & \mathbf{b}_2 \cdot \mathbf{c}_2 & \cdots &  \mathbf{b}_2 \cdot \mathbf{c}_{\tilde \nu} \\
                 \vdots & \vdots &  \ddots  & \vdots  \\ 
\mathbf{b}_\gamma \cdot \mathbf{c}_1 & \mathbf{b}_\gamma \cdot \mathbf{c}_2 & \cdots &  \mathbf{b}_\gamma \cdot \mathbf{c}_{\tilde \nu} ]
    }_{\mathbf{G}}
    \mqty[  \mathbf{d}_1^T \\  \mathbf{d}_2^T \\ \vdots \\  \mathbf{d}_{\tilde \nu}^T ]
\end{align}
For any generic $\dD(k)$ and generic $\omega \in \CC$, the singular values of $\mathbf{G}$ will be nonzero. (By \emph{generic}, we mean that the only exceptions, if any, constitute a measure zero sets). In this case, we have $\rank \mathbf{B} \cdot \mathbf{H} = \min(\tilde \nu, \gamma)$. Hence by the rank nullity theorem, $\dim \ker \mathbf{B} \cdot \mathbf{H} =\tilde \nu - \min(\tilde \nu,\gamma) $, which is the index theorem given in the methods. 
For solutions on the domain $x  \in ( - \infty, 0]$, a similar derivation applies. The difference being that $f(k)$ has $d-\tilde \nu (\omega)$ zeros with $\Im(k) < 0$.

We now provide two examples physically relevant independent homogeneous boundary conditions. Stress-free boundary conditions can be formulated by choosing $\mathbf{B}$ as follows: 
\begin{align}
   \mathbf{B} = \mqty[ 0 & 0 & C_{11} & C_{12}  \\ 0 & 0 & C_{21} & C_{22} ] 
\end{align}
where $\mathbf C$ is the constitutive matrix. For motion-free boundary  conditions, one may write
\begin{align}
    \mathbf{B} = \mqty[ 1 & 0 & 0 & 0 \\ 0 & 1 & 0 & 0] \label{eq:motfree}
\end{align}
which imposes that $\tilde p_h = \tilde p_\varphi =0$ at the boundary, i.e. there is no rotation or translation in time at the boundary.  To impose displacement free boundary conditions, one can reformulate the linear differential equation such that $\Psib=(\tilde p_h , \tilde p_\varphi, h, \varphi )$ instead of $\Psib=(\tilde p_h , \tilde p_\varphi, \tilde s, \tilde b)$.

\subsection{Boundary conditions in discrete models}

Here we derive and interpret $\nu(\omega)$ in discrete settings, such as discrete models or finite element simulations. We refer the reader to Refs.~\cite{trefethen2020spectra, Lee2019}, and references therein, for additional discussion. For concreteness, we will illustrate each step with the discrete model in Section IIF of the main text. To begin, suppose that the $i$th unit cell of the discrete model is described by an $n$-component vector $\Psib_i$. For the example at hand, we have $n=4$ and $\Psib_i = (p_i, l_i, h_i, \varphi_i)$ where $p_i = m \dot h_i$ is the vertical linear momentum and $ l_i = J \dot \varphi_i $ is the angular momentum of each unit cell about its center of mass. The linear equations of motion for an infinite system take the form:
\begin{align}
    i \partial_t \mqty[  \vdots  \\ \Psib_{i-1} \\ \Psib_i \\ \Psib_{i+1} \\ \vdots  ] =  
   \underbrace{\begin{bmatrix}  \ddots  &       &   \ddots    &  \ddots  &         \ddots      &        & \ddots      & \\
                       & \DD_l&   \cdots   &   \DD_{-1}       & \DD_0          & \DD_1       &  \cdots         & \DD_r & 0 &  \\
                       &    0     &\DD_l   & \cdots   & \DD_{-1}       & \DD_0     & \DD_1     &  \cdots & \DD_r & \\
                       &     &      & \ddots      &        & \ddots       & \ddots       &          &   & \ddots  
                       \end{bmatrix} }_{\DD}
    \mqty[ \vdots \\ \Psib_{i-1} \\ \Psib_i \\ \Psib_{i+1} \\ \vdots  ]
\end{align}
where the matrix $\DD$ is an infinite matrix formally known as an \emph{Laurent operator}. Each $n \times n$ matrix $\DD_i$ lies in the $i$th off diagonal, with $i$ ranging from $l <0$ to $r>0$. For the system at hand, we have $l=r=-1$, and the equations of motion in Eqs.~(21-22) of the main text take the form
\begin{align}
    i \partial_t \mqty[ \vdots \\ \Psib_{i-1} \\ \Psib_i \\ \Psib_{i+1} \\ \vdots  ] =  
   \mqty[  \ddots &  \ddots  &          &           & \\
                       \ddots & \DD_0        & \DD_1        &  0        & \\
                              & \DD_{-1}       & \DD_0        & \DD_1         & \\
                              & 0        & \DD_{-1}        & \DD_0         & \ddots  \\
                              &          &          &  \ddots   & \ddots ] 
    \mqty[ \vdots \\ \Psib_{i-1} \\ \Psib_i \\ \Psib_{i+1} \\ \vdots  ]
\end{align}
with 
\begin{align}
    \DD_0 =&i \mqty[ 0  & 0  & -2k_\mu  &  -k_\mu L - 2 p  \\
                 0  & 0  &  -k_\mu L   & -k_\mu L^2 -2 \kappa_B - 2 pL  \\
                 \frac1m  & 0  &  0   &  0        \\
                 0  & \frac1J  &  0   &  0 ] \\
    \DD_1 =& i\mqty[ 0 & 0 & k_\mu  &  p  \\
                 0 & 0 &  k_\mu  L & \kappa_B + p L \\
                 0 & 0 & 0 & 0  \\
                 0 & 0 & 0 & 0]\\
    \DD_{-1}=&i \mqty[ 0 & 0 & k_\mu  & Lk_\mu + p \\
                    0 & 0 &  0  & \kappa_B    \\
                    0 & 0 & 0 & 0 ]
\end{align}
For an infinite system, one is interested in the spectrum of $\DD$. To find the eigenvalues, one can perform a Fourier transform to obtain the symbol 
\begin{align}
    \DD(\lambda) = \sum_{i=l}^r \DD_i \lambda^i
\end{align}
where $\lambda = e^{ik L }$ with $k$ being the wavenumber and $L$ being the lattice spacing. 
We are then interested in the solutions to the equation:
\begin{align}
    F(\lambda) \equiv \det[ \DD(\lambda) - \omega ] =0
\end{align}
For the example at hand, 
\begin{align}
    F(\lambda) =& \frac{k_\mu  \kappa_B }{J m} \lambda^2 - \frac{ 4 k_\mu \kappa_B - (J k_\mu + L m p + m \kappa_B ) \omega^2 }{Jm} \lambda \nonumber \\
    &+ \frac{ 2 k_\mu L p + 6 k_\mu \kappa_B - ( 2 J k_\mu + k_\mu L^2 m + 2 L m p + 2 m \kappa_B  ) \omega^2 + J m \omega^4 }{Jm}  \nonumber \\
    &- \frac{ 4 k_\mu \kappa_B - (J k_\mu + m \kappa_B ) \omega^2}{ Jm } \frac1\lambda + \frac{k_\mu \kappa_B }{J m } \frac1{\lambda^2}
\end{align}
For the semi-infinite system bounded on the left, then we are interested in the solutions of the following equation:

\begin{align}
    i \partial_t 
       \mqty[ \Psib_0 \\ \Psib_{1} \\ \Psib_2 \\ \vdots  ]=
   \underbrace{\begin{bmatrix} 
     \DD_0          & \DD_1       &  \cdots         & \DD_r & 0 &  \\
      \DD_{-1}       & \DD_0          & \DD_1       &  \cdots         & \DD_r & 0 &  \\
        \vdots & \ddots & \ddots & \ddots & & \ddots & \ddots   \\
        \DD_l   & \cdots   &   \DD_{-1}       & \DD_0          & \DD_1       &  \cdots         & \DD_r & 0 &  \\
                    0   & \DD_l&   \cdots   &   \DD_{-1}       & \DD_0          & D_1       &  \cdots         & \DD_r & 0 &  \\
                            &      & \ddots      &        & \ddots       & \ddots       &  \ddots         &   & \ddots 
                       \end{bmatrix} }_{\tilde \DD_L}
    \mqty[ \Psib_0 \\ \Psib_{1} \\ \Psib_2 \\ \vdots  ]
\end{align}
Here $\tilde \DD_L$ is a semi-infinite matrix known as a \emph{Toeplitz operator}, which is the truncation of a Laurent operator. We seek to determine the number of eigenvectors of $\tilde \DD_L$ with eigenvalue $\omega$, or equivalently $\dim \ker  [ \tilde \DD_L - \omega ]$. Notice that $\ker [\tilde \DD_L -\omega] $ is isomporphic to the subspace of $\ker [\DD- \omega] $ that obeys the constraint:
\begin{align}
    0 = \underbrace{
    \begin{bmatrix}  
     \DD_{-1} & 0      & 0  & 0  \\
     \DD_{-2} & \DD_{-1} & 0  & 0 \\
     \vdots & \ddots & \ddots & 0 \\
     \DD_{l}  & \cdots & \DD_{-2} & \DD_{-1} 
    \end{bmatrix} }_{\Mb_L}
    \begin{bmatrix}
    \Psi_{l} \\ \Psi_{l+1} \\ \vdots \\ \Psi_{-1}
    \end{bmatrix} \label{eq:constraint}
\end{align}
The constraint in Eq.~(\ref{eq:constraint}) can be interpreted as the boundary conditions implied by the truncation of the Laurent operator. The number of independent boundary conditions is given by $\gamma_L = \rank \Mb_L$. For the discrete model of the beam, the constraint reads $\DD_{-1} \cdot \Psib_{-1} =0$, or equivalently $ h_{-1} =0$ and $\varphi_{-1} =0$, which are displacement and rotation free boundaries. Notice that in general, the form of $\Mb_L$ depends on how the continuum equation is discretized. 
More generally, one can show that $\rank \Mb = \deg F(1/\lambda)$, where $\deg$ denotes the highest power appearing in the polynomial. For the discrete model of the odd micropolar beam, we have $\gamma_L=2$.

Next we consider the integral
\begin{align}
    \nu(\omega) = \frac1{2 \pi i }\oint_{S^1} \dv{\lambda } \log F(\lambda) \dd \lambda 
\end{align}
where $S^1$ is the unit circle. By Cauchy's argument principle, $\nu(\omega)$ counts with multiplicity the number of zeros of minus the number of poles of $F(\lambda)$ with $\abs{\lambda} \le 1$ (or equivalently $\Im(k) > 0$). Each of the zeros represents and eigenmode of $\DD(\lambda)$ there are exactly $\gamma_L$ poles at $\lambda =0$ corresponding to the $\gamma_L$ constraints imposed by the matrix $\Mb$. Next, supposing $\nu + \gamma_L > 0$, let $\lambda_a$ and $\Psib^a_i$, with $a=1,2, \dots, \nu(\omega)+ \gamma_L$ be the roots and eigenvectors, respectively. Then we may define the matrix
\begin{align}
    H = \mqty[ \Psib_{l}^1 & \cdots & \Psib^{\nu + \gamma_L }_l  \\
    \vdots &  & \vdots    \\
    \Psib_{-1}^1 & \cdots & \Psib^{\nu+\gamma_L}_{-1}
    ] 
\end{align}
We seek to determine $\dim \ker \Mb_L \cdot \Hb$. Notice that $\rank \Hb = \min( \nu+\gamma_L, n \abs{l}  )$ and $\rank \Mb = \gamma_L < n \abs{l}$. Therefore, for a generic $\DD(\lambda)$ and a generic $\omega \in \CC$ we have $\rank \Mb_L \cdot \Hb =  \min(\gamma_L, \nu+ \gamma_L) $. Hence, by the rank-nullity theorem, $\dim \ker \Mb_L \cdot \Hb = \max( 0, \nu ) $. Thus there will generically be $\max[\nu(\omega), 0] $ eigenmodes of $\tilde \DD_L$ at frequency $\omega$ localized to the left boundary. Likewise, to study eigenmodes localized to the right, one obtains $\tilde \DD_R$ by truncating $\DD$ in the opposite direction. In this case, one
is interested in the constraint matrix:
\begin{align}
    \Mb_R = \mqty[ \DD_l & \DD_2 & \cdots & \DD_1 \\
                 \vdots &  &\reflectbox{$\ddots$}  &  \\
                 \DD_2 &  \DD_1 \\ 
                 \DD_1 & & 
    ]
\end{align}
For $\Mb_R$, we have $\gamma_L = \rank \Mb_R = \deg F(\lambda)$. Repeating the argument above, for a generic $\DD(\lambda)$ and $\omega \in \CC$, one has $\rank \ker \Mb_R \cdot \Hb = \max[0, - \nu(\omega) ] $ and thus there will be  $\max[0, - \nu(\omega) ]$ eigenmodes of $\tilde \DD_R $ localized to the right boundary. 

Finally, we discuss the application to numerical eigenmode solvers. The COMSOL simulations impose a finite element mesh on a single unit cell, whose points form the content of the vector $\Psib_i$ that is governed by an $n \times n $ matrix $\DD(\lambda)$, now with a much larger $n$. When solving for eigenmodes at a given wavenumber $k$,  COMSOL requires that the state of the right boundary is equal to the state of the left boundary of the mesh multiplied by a phase $e^{i k L}$. This is equivalent to implementing surface coupling between nearest neighbor unit cells, corresponding to $r = -l =1$. For this surface coupling, the constraint matrices $\Mb_R$ and $\Mb_L$ effectively impose motion-free boundary conditions. In the continuum, motion-free boundaries are captured by Eq.~(\ref{eq:motfree}), for which $\gamma =2$. Hence, for the data in Fig.~3 and Fig.~M1, one should compare $\nu(\omega) = \tilde \nu(\omega)-2$.

\subsection{Explicit calculation of eigenmodes}
 
Here we provide explicit calculations of eigenmodes in the continuum theory. The starting place is the wave equation in Eq.~(23) of the main text, repeated here:
\begin{align}
{\omega}
{\begin{bmatrix}
\tilde p_h \\ \tilde p_\varphi \\ \tilde s \\ \tilde b 
\end{bmatrix}}{=} 
\underbrace{
{\omega_1}{\begin{bmatrix}
 0 & 0 & - k \ell_1 & - \tilde P \ell_1 k \\
0& 0 & i & i \tilde P  - k  \ell_2 \\
-k \ell_1 & -i & 0 & 0 \\
0 & -k \ell_2 & 0 & 0
\end{bmatrix}} }_{\dD (k)}
{\begin{bmatrix}
\tilde p_h \\ \tilde p_\varphi \\ \tilde s \\ \tilde b 
\end{bmatrix}} \label{eq:dmat2}
\end{align}
The secular equation is given by 
\begin{align}
        0= \tilde \omega^4 -  \qty[ 1- i \tilde P k \ell_2 +k^2 (\ell_1^2 + \ell_2^2) ] \tilde \omega^2 + k^4 \ell_1^2 \ell_2^2 \label{eq:secsi} 
\end{align}
First notice that when $P=0$ and to leading order in $k=0$, the eigensystem of $\dD(k)$ takes the following form:
\begin{align}
    \begin{bmatrix}
\tilde p_h \\ \tilde p_\varphi \\ \tilde s \\ \tilde b  
\end{bmatrix}
= 
    \begin{bmatrix}
\pm i \\ k \ell_1 \\ \pm i k \ell_2 \\ 1  
\end{bmatrix}
\, \, \text{ with } \omega = \pm k^2 \ell_1 \ell_2 \omega_1
\end{align}
which represents a bending dominated goldstone mode. The second two modes are shearing dominated gaped modes:
\begin{align}
    \begin{bmatrix}
\tilde p_h \\ \tilde p_\varphi \\ \tilde s \\ \tilde b  
\end{bmatrix}
= 
    \begin{bmatrix}
\pm i k \ell_1  \\ \pm 1  \\ i   \\ k \ell_2   
\end{bmatrix}
\, \, \text{ with } \omega = \pm \qty(1-  k^2 \frac{ \ell_1^2 + \ell_2^2 }2 ) \omega_1
\end{align}

When $P>0$, we proceed numerically. Let us use numerical values representative of the experiment, $\ell_1 = 10^{-3} \si{m}$, $\ell_2 = 10^{-2} \si{m}$, $\tilde P=1$, and $\omega_1 = 10^5 \si{Hz}$. We will consider the modes at two select frequencies: $\omega_A = 10^3 \si{Hz}$ and  $\omega_B = 10^6 \si{Hz} $. Notice that $\omega_A \ll \omega_1 \ll \omega_B$ and that the degree of Eq.~(\ref{eq:secsi}), treated as a polynomial in $k$, is $d =4$. From the location of $\omega_A$ and $\omega_B$ in the complex plane, we have $\nu(\omega_A) =1 $ and $\nu(\omega_B) =2$, see Fig.~\ref{fig:modes}a-b. Therefore, for $\omega_A$, we expect one solution to Eq.~(\ref{eq:secsi}) left-localized solution ($\Im(k) > 0$) and $d-\nu(\omega_A) =3$ right-localized solutions ($\Im(k) < 0$), see Fig.~\ref{fig:modes}c-d. We plot the corresponding eigenmode of $D(k)$ for each of these values, and we find that there is stronger mixing between bending and shearing than in the passive case.

To illustrate the meaning of $\tilde \nu(\omega_A)=1$, suppose we impose two boundary conditions ($\gamma=2$) such as $\tilde s = \tilde b =0$ at $x =0$. Given that there is only one left-localized mode, the two boundary conditions cannot be satisfied for a system with a boundary on the left. However, there are three right-localized modes so there is one nonzero linear combination of right localized modes that satisfy the boundary conditions. We can repeat the calculation now for $\omega_B$, for which we have $\tilde \nu(\omega_B) =2$. Hence we expect $\tilde \nu(\omega_B)=2$ left localized modes and $d-\tilde \nu(\omega_B)=2$ right localized modes before considering boundary conditions. Once we impose $\gamma=2$ boundary conditions (e.g.  $\tilde s = \tilde b=0$), one finds that it is not possible to formulate a linear combination for a nonzero mode at either boundary.

\subsection{Role of additional vibrational modes} 
As shown in Fig.~M1 of the main text, the continuum theory predicts a total of four vibrational modes: two flexural goldstone modes and two gapped, shear dominated modes. However, at high frequencies, comparable to $\omega_1$, the continuum theory is not expected to self-consistently apply. Hence, the shear dominated modes should not necessarily be thought of as a physical prediction made by the continuum theory. Moreover, there is a second simplifications invoked when using this continuum theory. The physical beam has a total of four goldstone modes: a flexural mode, a torsional mode, an out-of-plane deformation, and a longitudinal mode.  Our ability to apply the continuum theory to the physical beam rests on two prerequisites. First the experiments probe frequencies on the order of $\lesssim 10 \si{kHz}$, well below $\omega_1 \approx 100 \si{kHz}$. Second, based on the construction of the piezoelectric feedback, we assume that at low frequencies, only the flexural mode significantly couples to the electronic feedback. To validate these assumptions, in Fig.~\ref{fig:high_freq} we use COMSOL to compute the lowest 15 vibrational modes for $P>0$. In panel a, the real part of the first 15 modes are plotted as a function of $k$, and the modes associated with selected branches are shown in panels b-d. (Modes from every branch are shown Fig.~\ref{fig:numerical_modes}.) In panels e-f, we plot the spectrum in the complex plane. Frequencies with negligible imaginary parts are indicated by grey dots, and the remaining frequencies are colored by their wave number $k$. As can be seen, for frequencies less than $\approx 7 \si{kHz}$, only the flexural modes (shown also in Fig.~3 of the main text) have a significant imaginary part. This validates our assumption that that the piezoelectric feedback primarily couples to the flexural modes at low frequencies. 

One can ask how the presence of additional vibrational modes affects the computation of the winding number $\nu(\omega)$ in Eq.~(30) of the main text. As shown in Fig.~\ref{fig:high_freq}e, only the flexural mode as a significant imaginary part at low frequencies, and so only flexural modes will be significantly localized in the frequency range of interest. Additionally, one can ask how to reconcile the continuum equation (Eq. M5) with the fact that the shear dominated band contributes to $\nu(\omega)$ despite being an unphysical artifact of the theory. The answer is that the physically relevant piece of information is not necessarily the absolute value of $\nu(\omega)$, but rather the relative value $\nu(\omega')-\nu(\omega'')$ for any two given frequencies $\omega'$ and $\omega''$.  Suppose, for example, there are $n$ left localized modes at $\omega'$ with a given set of boundary conditions. The absolute value of $\nu(\omega')$ and the way in which the boundary conditions are quantified will depend on the details of the theory. However, given only the value of $\nu(\omega') - \nu(\omega'')$, one can conclude that there will be $n- \nu(\omega') + \nu(\omega'') $ left localized modes at $\omega''$ for the same set of boundary conditions. Finally, we note that to compute the relative value $\nu(\omega')-\nu(\omega'')$, one need only draw a line in the complex plane from $\omega''$ to $\omega'$ and count the number of signed crossings with the periodic boundary spectrum. Hence, the value $\nu(\omega')-\nu(\omega'')$ only depends on accurately resolving the spectrum in the frequency range of interest, and not on features of the spectrum outside the range of validity of the theory.

\subsection{Simulations of quasistatic energy cycle and finite frequency efficiency}

In Fig.~2a-b., we verify Eq.~(20) via finite element simulations. We  directly compute the work done by a single unit cell of the material that undergoes a path of deformations via fully electromechanically coupled numerical simulations in COMSOL Multiphysics. We enforce displacement boundary conditions on the side boundaries of a metamaterial unit cell. We define the bending curvature $\kappa = \partial_x \varphi$ to be the relative angle of the two terminating cross sections divided by the unit cell's length $L$, and $\gamma = \partial_x h$ to be the relative vertical displacements of the two cross sections divided by the unit cell's length. In the calculations, geometric and material parameters of the metamaterial are given in Table S1 (See below). We take the beam on a rectangular path through deformation space with $\gamma_\text{max} = 9.375 \times 10^{-3}$ and $\kappa_\text{ max} = 1.0 \ m^{-1}$. We subdivide each leg of the path into approximately 150 subdivisions and perform a static analysis at each subdivision to determine the reaction forces $F_r$ and $F_l$ and moments $M_l$ and $M_r$ at the left and right boundaries, respectively. We then compute the work done on the $i$th step as:
\begin{align}
    W^i = L \qty[ (F^i_r -F^i_l)(\gamma^{i} - \gamma^{i-1}) + (M^i_r - M^i_l )(\kappa^{i} - \kappa^{i-1})]
\end{align}
where the index $i$ labels the step.

For the example shown in the main text, $P = 13.85\times 10^6 ~N/m$ and the volume of the unit cell is $V= 576\times10^{-9} ~m^3$. 
The continuum theory then predicts that the magnitude of the work done should be given by:
\begin{align}
\left|W_{\text{ theory}} \right|=P\kappa^{max}\gamma^{max}V=0.0748\left(J\right).
\end{align}
In simulation, we find:
\begin{align}
\left| W_{\text {sim}} \right| = \left | \sum_i  W^i \right | = 0.0756 \left( J \right),
\end{align}
We note that the continuum equations rely on the approximation of linearity, whereas the COMSOL simulations directly compute the underlying forces and moments based on the microscopic details.

\begin{figure}[h!]
    \centering
    \includegraphics[width=0.4\textwidth]{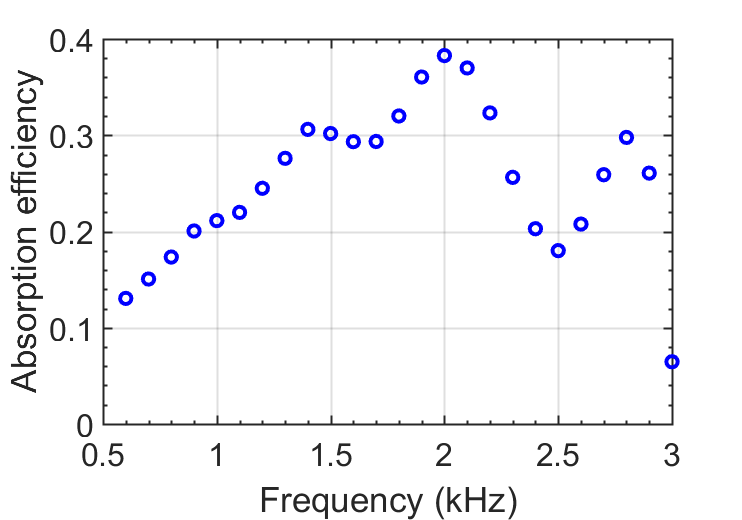}
    \caption{The absorption efficiency $\mathcal{E}$ as a function of frequency for an incident flexural wave.}
    \label{fig:eff}
\end{figure}

In addition to quasistatic strain controlled deformations, we also examine the efficiency of our metabeam at absorbing energy from finite frequency waves. To do so, we first perform numerical simulations matching the parameters in our experiments. From these numerical tests, we measure the mechanical energy flux from the left and right boundaries of a given unit cell (in this case the fifth unit cell). Denoting these energy fluxes by $F_L$ and $F_R$, respectively, we define the energy absorption efficiency of left-traveling waves by $\mathcal{E} = (F_R - F_L)/F_R$, where $\abs{F_R} >\abs{F_L}$. At 2 kHz, we find the peak absorption efficiency per unit cell is equal to $\mathcal{E} =0.38$. Supplementary Figure~\ref{fig:eff} shows this efficiency at frequencies from 0.6 to 3.0 kHz, where the waves propagating from the right to the left are attenuated.

\section{Experimental details} \label{expt}

\subsection{Material parameters}
\begin{figure}[ht]
    \centering
    \includegraphics{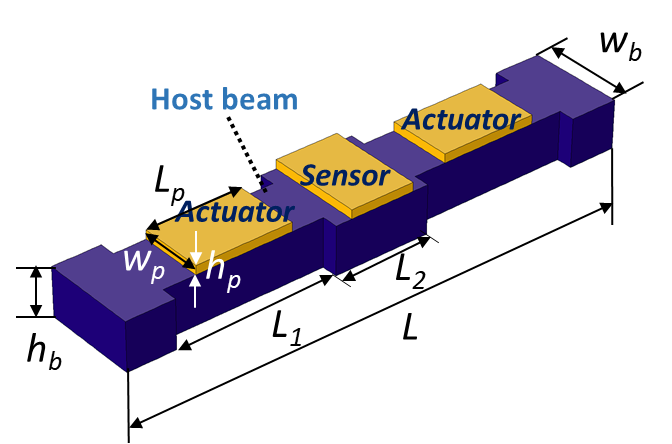}
     \caption{ Schematic of a unit cell.}
    \label{fig:unitcell}
\end{figure}

The active metamaterial displayed in main-text Fig.~1 utilizes piezoelectric patches integrated with electronic feedback~\cite{Chen2014jva,Bergamini2014,Chen2016, Yi2017,Alan2019,Chen2017,Wang2017,Casadei2012,Chen2016iop,Sugino2018,Lakes2012}. Supplementary Figure~\ref{fig:unitcell} provides a schematic of the odd micropolar metamaterial. Tables S1 and S2 contain the geometric and material parameters, respectively, used in the design.

\input{table1}
\input{table2}

\subsection{Electrical control system}
\begin{figure}[ht]
	\centering
	\includegraphics[width=\textwidth]{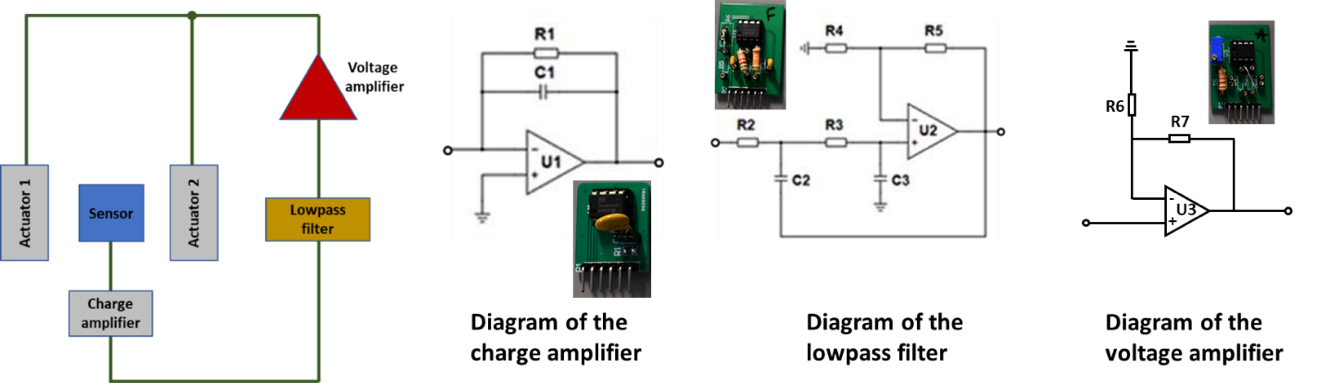}
	\caption{ The schematic of the electrical control system and circuit diagrams for its individual components. }
	\label{fig:electronics}
\end{figure}

Supplementary Figure~\ref{fig:electronics} shows a schematic of the electrical control system used in each unit cell. We use a standard non-inverting voltage amplifier, and all the electrical circuits were fabricated on printed circuit boards (PCBs). The parameters for the electrical components are listed in Table S3. The frequency dependence of the lowpass filter is shown in main text Eq.~(40) and depicted in Supplementary Figure~\ref{fig:lowpass}.

During experiments, we also need to consider stability conditions of the metamaterial. In particular, we find that small feedback effects emerge within individual unit cells associated with imperfections in fabrication. When $\abs{H_0}$ exceeds a critical value, the antisymmetric actuating voltages can no longer produce zero sensing signals in experiments and the system experiences an instability. This critical value is $\abs{H_0} \approx 6$ in our current system, but it is highly dependent on the fabrication details. Due to the self amplification of the beam, the critical value of $\abs{H_0}$ usually decreases as the number of unit cells increases.

\input{table3}

\begin{figure}[ht]
    \centering
    \includegraphics[width=0.6\textwidth]{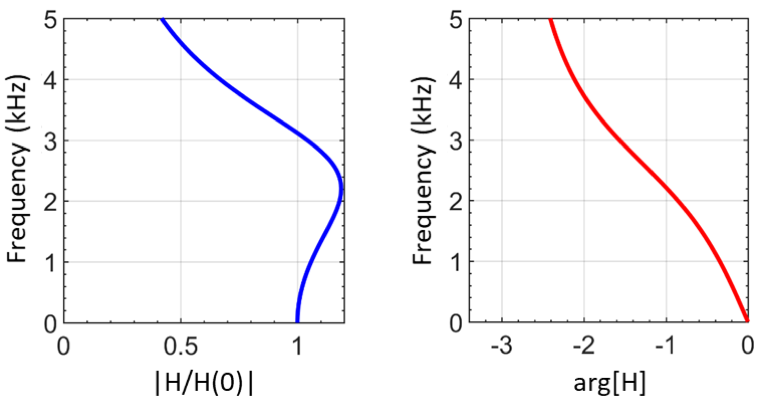}
    \caption{Frequency response of the lowpass filter with the transfer function $H(\omega)$. Left: Amplitude; Right: Phase angle.}
    \label{fig:lowpass}
\end{figure}

\section{Numerics} \label{numerics}

\subsection{Transfer matrix method for wave dispersion} \label{sec:TMM}

Here we describe the transfer matrix method~\cite{Chen2016iop} used to produce semi-analytical curves in Figures~3 and 6 in the main text. In this approach, the piezoelectric sensing patch is idealized as a point-like strain probe located in the middle each unit cell. This approximation is justified by the large ratio between the experimentally probed wavelengths and the length of the piezoelectric patch (see Supplementary Figure~\ref{fig:TMM}). Similarly, piezoelectric actuating patches are idealized as point sources that generate bending moments (see Supplementary Figure~\ref{fig:TMM}). With these approximations, the metamaterial unit cell can be divided into seven homogeneous beam sections, shown in Supplementary Figure~\ref{fig:TMM}. 

\begin{figure}[ht]
	\centering
	\includegraphics{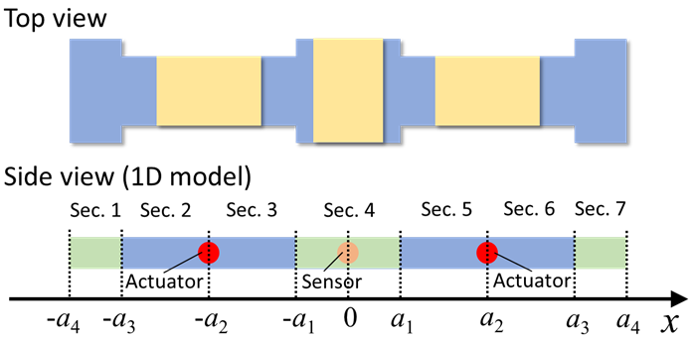}
	\caption{Schematic for the transfer matrix method.}
	\label{fig:TMM}
\end{figure}

For each section, we apply Timoshenko beam assumptions, yielding the following equation of motion for the $n$-th beam section:

\begin{align}
   B_n\frac{\partial^4w_n}{\partial x^4}+\left(\frac{B_n\rho_n\omega^2}{\mu_n}+I_n\omega^2\right)\frac{\partial^2w_n}{\partial x^2}+\frac{\left(I_n\omega^2-\mu_n\right)\rho_n\omega^2}{\mu_n}w_n=0,
\end{align}
where $w_n$, $B_n$, $\rho_n$, $\nu_n$, and $I_n$ denote transverse displacement, bending stiffness, mass density per unit volume, shear modulus and moment of inertia of the n-th homogeneous beam section, respectively. Furthermore, $\omega$ is the frequency associated to the Bloch mode $e^{-i\omega t}$. The solutions of $w_n$ read 
\begin{align}
w_n=\bar{A}_ne^{-ik_nx}+\bar{B}_ne^{ik_nx}+\bar{C}_ne^{-\hat{k}_nx}+\bar{D}_ne^{\hat{k}_nx},
\end{align}
where $k_n=-i\sqrt{\frac{-\alpha_n-\sqrt{\alpha_n^2-4B_n\beta_n}}{2B_n}}$ and ${\hat{k}}_n=\sqrt{\frac{-\alpha_n+\sqrt{\alpha_n^2-4B_n\beta_n}}{2B_n}}$ with $\alpha_n=\frac{\rho_n\omega^2B_n}{\mu_n}+I_n\omega^2$ and $\beta_n=\frac{\left(I_n\omega^2-\mu_n\right)\rho_n\omega^2}{\mu_n}$. In the design, beam sections 1, 4 and 7 possess the same material and geometric parameters. Hence, we may simply write $k_n =k_1$ and $\hat k_n =\hat k_1$ for these sections. Similarly, we may simply use $k_n=k_2$ and $\hat k_n =\hat k_2$ for sections 2, 3, 5 and 6. For beam sections 1, 4 and 7, we may summarize the forces and displacement in matrix form by:
\begin{align}
\mathbf{W}_n\left(x\right)=\mathbf{N}_1\left(x\right)\mathbf{A}_n, ~n=1,4,7 \label{Weq}
\end{align}
In Eq.~(\ref{Weq}),  $\mathbf{W}_n\left(x\right)=\left[\begin{matrix}w_n\left(x\right)&\varphi_n\left(x\right)&M_n\left(x\right)&F_n\left(x\right)\\\end{matrix}\right]^T$, where $\varphi_n\left(x\right)$, $M_n\left(x\right)$ and $F_n\left(x\right)$ are the rotational angle, bending moment and shear force in the $n$-th beam section. Furthermore, $\mathbf{A}_n=\left[\begin{matrix}
\bar{A}_n& \bar{B}_n& \bar{C}_n& \bar{D}_n
\end{matrix}\right]^T$. Lastly,  
\[  
\mathbf{N}_1\left(x\right)=\left[\begin{matrix}e^{-ik_1x}&e^{ik_1x}&e^{-{\hat{k}}_1x}&e^{{\hat{k}}_1x}\\\frac{-ik_1^2\mu_1+i\omega^2\rho_1}{k_1\mu_1}e^{-ik_1x}&\frac{ik_1^2\mu_1-i\omega^2\rho_1}{k_1\mu_1}e^{ik_1x}&\frac{-{\hat{k}}_1^2\mu_1-\omega^2\rho_1}{{\hat{k}}_1\mu_1}e^{-{\hat{k}}_1x}&\frac{{\hat{k}}_1^2\mu_1+\omega^2\rho_1}{{\hat{k}}_1\mu_1}e^{{\hat{k}}_1x}\\\frac{-k_1^2\mu_1+\omega^2\rho_1}{\mu_1}B_1e^{-ik_1x}&\frac{-k_1^2\mu_1+\omega^2\rho_1}{\mu_1}B_1e^{ik_1x}&\frac{{\hat{k}}_1^2\mu_1+\omega^2\rho_1}{\mu_1}B_1e^{-{\hat{k}}_1x}&\frac{{\hat{k}}_1^2\mu_1+\omega^2\rho_1}{\mu_1}B_1e^{{\hat{k}}_1x}\\\frac{-i\omega^2\rho_1S_1}{k_1}e^{-ik_1x}&\frac{i\omega^2\rho_1S_1}{k_1}e^{ik_1x}&\frac{\omega^2\rho_1S_1}{{\hat{k}}_1}e^{-{\hat{k}}_1x}&\frac{-\omega^2\rho_1S_1}{{\hat{k}}_1}e^{{\hat{k}}_1x}\\\end{matrix}\right]
\]
where $S_n$ is the area of the cross-section of the $n$-th beam. Similarly, for beam sections 2, 3, 5 and 6, Eq.~(\ref{Weq}) reads
\begin{align}
\mathbf{W}_n\left(x\right)=\mathbf{N}_2\left(x\right)\mathbf{A}_n,~n=2,3,5,6
\end{align}
where $N_2(x)$ is obtained by replacing the index ``1" in $N_1(x)$ by ``2”.

The effective point source vector generated by the left actuator can be written as 
\begin{align}
\mathbf{G}=\mathbf{H}\mathbf{A_4}
\end{align}
where $\mathbf{G}=\left[\begin{matrix}0&0&M_a&0\end{matrix}\right]^T $  with $M_a$ being the effective bending moment produced by the actuator, and 
$$\mathbf{H}=\left[\begin{matrix}0&0&0&0\\0&0&0&0\\H\kappa_a\kappa_s&H\kappa_a\kappa_s&H\kappa_a\kappa_s&H\kappa_a\kappa_s\\0&0&0&0\\\end{matrix}\right]$$
where $\kappa_a$ and $\kappa_s$ denote electromechanical coupling coefficients of the piezoelectric actuator and sensor, respectively, which are retrieved from finite-element simulations.

Next, we impose continuity conditions on the transverse displacement, rotational angle, bending moment and shear force at the section boundaries ($x = -a_3, -a_2, -a_1, a_1, a_2$ and $a_3$) to obtain:

\begin{align}
\begin{matrix*}[l]
\mathbf{N}_1\left(-a_3\right)\mathbf{A}_1=\mathbf{N}_2\left(-a_3\right)\mathbf{A}_2,\\
\mathbf{N}_2\left(-a_2\right)\mathbf{A}_2=\mathbf{N}_2\left(-a_2\right)\mathbf{A}_3+\mathbf{H}\mathbf{A}_4,\\
\mathbf{N}_2\left(-a_1\right)\mathbf{A}_3=\mathbf{N}_1\left(-a_1\right)\mathbf{A}_4,\\
\mathbf{N}_1\left(a_1\right)\mathbf{A}_4=\mathbf{N}_2\left(a_1\right)\mathbf{A}_5,\\
\mathbf{N}_2\left(a_2\right)\mathbf{A}_5=\mathbf{N}_2\left(a_2\right)\mathbf{A}_6-\mathbf{H}\mathbf{A}_4,\\
\mathbf{N}_2\left(a_3\right)\mathbf{A}_6=\mathbf{N}_1\left(a_3\right)\mathbf{A}_7.\\
\label{Neq}
\end{matrix*}
\end{align}
Eq. (\ref{Neq}) can be expressed as
\begin{align}
\mathbf{A}_7=\mathbf{T}\mathbf{A}_1, \label{trans}
\end{align}
where the transfer matrix $\mathbf{T}$ is given by
\begin{align}
\mathbf{T}=\mathbf{N}_1^{-1}\left(a_3\right)\mathbf{N}_2\left(a_3\right)\mathbf{N}_2^{-1}\left(a_2\right)\left[\mathbf{N}_2\left(a_2\right)+\mathbf{H}\mathbf{N}_1^{-1}\left(a_1\right)\mathbf{N}_2\left(a_1\right)\right]\mathbf{N}_2^{-1}\left(a_1\right)\mathbf{N}_1\left(a_1\right)\mathbf{N}_1^{-1}\left(-a_1\right) \times \\
\mathbf{N}_2\left(-a_1\right)\left[\mathbf{N}_2\left(-a_2\right)+\mathbf{H}\mathbf{N}_1^{-1}\left(-a_1\right)\mathbf{N}_2\left(-a_1\right)\right]^{-1}\mathbf{N}_2\left(-a_2\right)\mathbf{N}_2^{-1}\left(-a_3\right)\mathbf{N}_1\left(-a_3\right).\nonumber 
\end{align}
Applying Bloch theorem on the left and right edges of the unit cell gives
\begin{align}
\mathbf{N}_1\left(-a_4\right)\mathbf{A}_1=e^{ikL}\mathbf{N}_1\left(a_4\right)\mathbf{A}_7. \label{bound}
\end{align}
Combining Eqs. (\ref{trans}) and (\ref{bound}), one can derive
\begin{align}
e^{ikL}\left[\mathbf{N}_1^{-1}\left(-a_4\right)\mathbf{N}_1\left(a_4\right)\mathbf{T}\right]\mathbf{A}_1=\mathbf{A}_1. \label{evp}
\end{align}
Solving the eigenvalue problem in Eq. (\ref{evp}) for imposed frequencies, one can obtain the corresponding complex wavenumbers.

\section{Movies}

\noindent {\bf Movie S1. Unidirectional amplification.} Experimentally measured transverse velocity wave field in response to excitation on the left. The incident wave is a tone burst signal centred at 2 kHz. 

\

\noindent {\bf Movie S2. Unidirectional attenuation.} Experimentally measured transverse velocity wave field in response to excitation on the right. The incident wave is a tone burst signal centered at 2 kHz.

\begin{figure}
    \centering
    \includegraphics{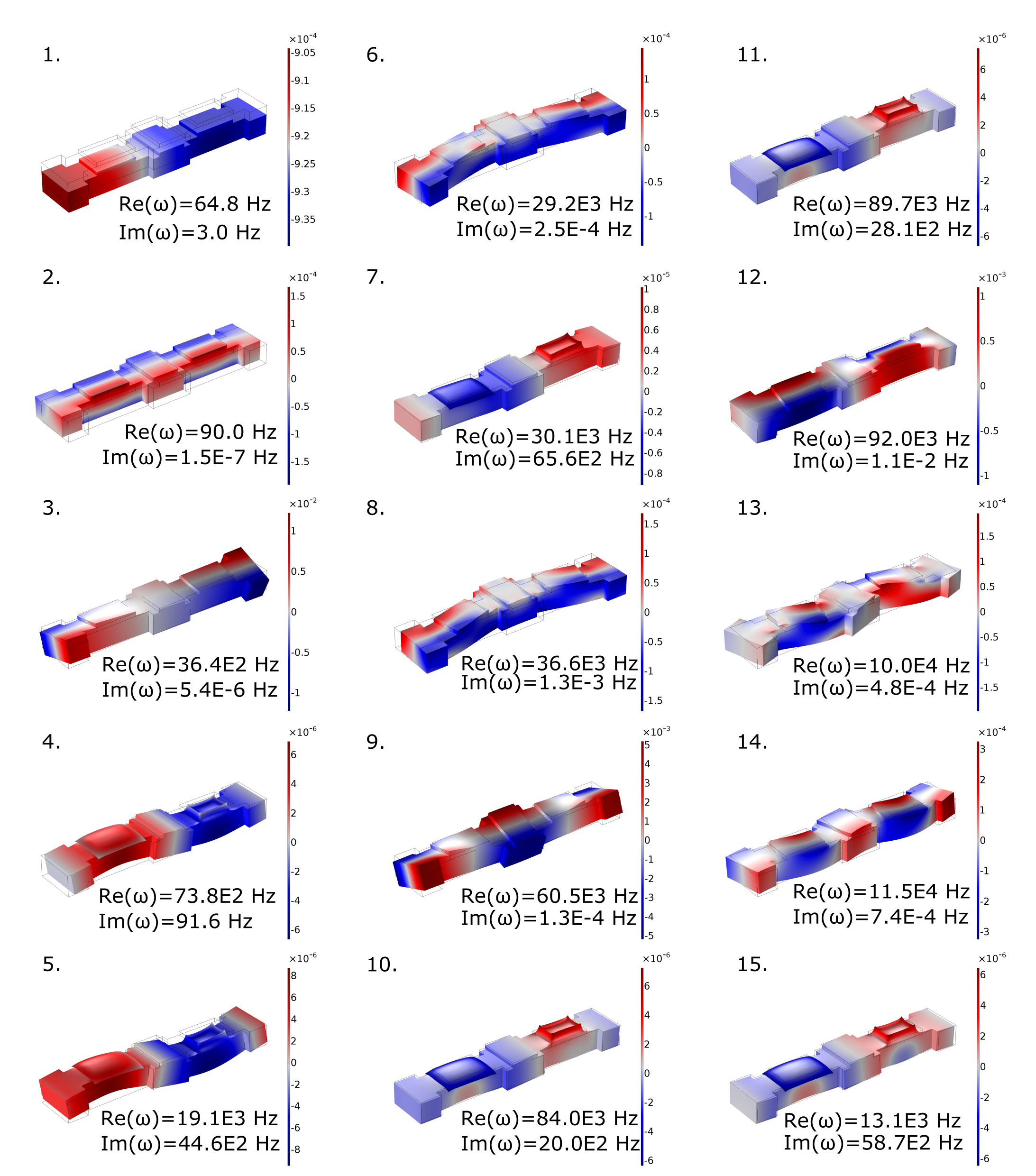}
    \caption{The lowest 16 modes corresponding to bands shown in Fig.~\ref{fig:high_freq}. The wavenumber is taken to be $kL =0.314$, where $L$ is the lattice spacing. The odd micropolar modulus is $P =3 \Pi$. The color bar denotes the $z$ (vertical) component of the displacement field.  }
    \label{fig:numerical_modes}
\end{figure}

\end{document}

%% file: table1.tex
\begin{table}[ht!]
\centering
\caption{ Geometric parameters of the odd micropolar metamaterial.}
\label{table1}
\begin{tabular}{clllllll|}
\hline \hline
\(L\)                      & 32  mm                    & \(h_b\)                                 & 3 mm                     & \(w_h\)                               & 6 mm         \\ 
\(L_p\)                  &  6 mm                        & {\(h_{p}\)}                       &  0.55 mm                     & \(w_{p}\)                           & 4 mm         \\ 
\(L_1\)                    & 10 mm                         & \(L_2\)                              & 6 mm                      &                              &           \\ \hline \hline
\end{tabular}
\end{table}

%% file: table2.tex
\begin{table}[ht!]
\centering
\label{table2}
\caption{ Material properties of the odd micropolar metamaterial.}
\begin{tabular}{llllll}
\hline\hline
\multicolumn{6}{l}{Material properties (Steel)}                                                             \\ \hline
\(E_b\)                       & 210.0 \(GPa\)           & \(G_b\)                       & 80.8 \(GPa\)      & \(\rho_b\)   & \(7800.0 ~kg  ~m^{-3}\)   \\ \hline
\multicolumn{6}{l}{Material properties (PZT 5J)}                                                                      \\ \hline
\(s_{11}^E\) & \(16.2\times 10^{-12} \,m^2N^{-1}\)  & \(d_{33}\)                   & \multicolumn{3}{l}{\(5.93 \times 10^{-10} \,C N^{-1}\)}   \\
\(s_{33}^E\) & \(20.7  \times 10^{-12}\, m^2 N^{-1}\)                     & \(d_{31}\)                    & \multicolumn{3}{l}{\(-2.74 \times 10^{-10} \,C N^{-1}\)}  \\
\(s_{44}^E\) & \(47  \times 10^{-12} \,m^2 N^{-1}\)                     & \(d_{15}\)                    & \multicolumn{3}{l}{\(7.41 \times 10^{-10} \,C N^{-1}\)}   \\
\(s_{12}^E\) & \(-4.54  \times 10^{-12} \,m^2 N^{-1}\)                  & \(\varepsilon_{33}^S\)                & \multicolumn{3}{l}{\(1433.6 \,\varepsilon_0\)}          \\
\(s_{13}^E\) & \(-5.9  \times 10^{-12}\, m^2 N^{-1}\)                  & \(\varepsilon_{11}^S\)                & \multicolumn{3}{l}{\(1704.4 \,\varepsilon_0\)}          \\ 
\(\rho_p\)      & \(7700.0 \,kg \,m^{-3}\)                                 & \(\varepsilon_{0}\)                       & \multicolumn{3}{l}{\(8.842  \times 10^{-12} \,C m V^{-1}\)} \\ \hline\hline
\end{tabular}
\end{table}

%% file: table3.tex
\begin{table}[ht!]
\centering
\caption{ Circuit component parameters.}
\label{table1}
\begin{tabular}{clllllll}
\hline \hline
R1                    & 1 M$\Omega$                   & R2                                & 9.09 k$\Omega$                     & R3                              &15.74 k$\Omega$        \\ 
C1                 &   1 nF                         & C2                       &  10 nF                      & C3                           & 2.2 nF        \\ 
Op-amp                   & OPA445                         &                               &                      &                              &           \\ \hline \hline
\end{tabular}
\end{table}